\documentclass[prd,aps,a4paper,twocolumn,eqsecnum,nofootinbib,floatfix]{revtex4}  %

\usepackage{amsmath,amsfonts,amssymb}

\newcommand{\beq}{\begin{equation}}
\newcommand{\eeq}{\end{equation}}
\newcommand{\bea}{\begin{eqnarray}}
\newcommand{\eea}{\end{eqnarray}}

\begin{document}

\title{Multipolar invariants and the eccentricity enhancement function parametrization of gravitational radiation}

\author{Donato Bini$^{1,2}$, Andrea Geralico$^1$}
  \affiliation{
$^1$Istituto per le Applicazioni del Calcolo ``M. Picone,'' CNR, I-00185 Rome, Italy\\
$^2$INFN, Sezione di Roma Tre, I-00146 Rome, Italy\\
}

\date{\today}

\begin{abstract}
Gravitational radiation can be decomposed as an infinite sum of radiative multipole moments, which parametrize the waveform at infinity. The multipolar-post-Minkowskian formalism provides a connection between these multipoles and the source multipole moments, known as explicit integrals over the matter source.
The gravitational wave energy, angular momentum and linear momentum fluxes are then expressed as multipolar expansions containing certain combinations of the source moments.
We compute several gauge-invariant quantities as \lq\lq building blocks" entering the multipolar expansion of both radiated energy and angular momentum at the 2.5 post-Newtonian (PN) level of accuracy in the case of hyperboliclike motion, by completing previous studies through the calculation of tail effects up to the fractional 1PN order. 
We express such multipolar invariants in terms of certain eccentricity enhancement factor functions, which are the counterpart of the well known enhancement functions already introduced in the literature for ellipticlike motion. 
Finally, we use the complete 2.5PN-accurate averaged energy and angular momentum fluxes to study the associated adiabatic evolution of orbital elements under gravitational radiation reaction.
\end{abstract}

\maketitle

\section{Introduction}

High-accuracy comparisons between the post-Newtonian (PN) predictions and the numerically-generated waveforms for gravitational waves emitted by inspiralling compact binaries need an even more precise description of both the orbital dynamics and the radiation field. 
The latter is decomposed into multipole moments according to the multipolar-post-Minkowskian (MPM) formalism, which relates them to the source moments \cite{Blanchet:1985sp,Blanchet:1989ki,Damour:1990ji,Blanchet:1998in,Poujade:2001ie}.
As a result, the radiated energy, angular momentum and linear momentum are expressed in terms of certain combinations of these moments, whose orbital averages provide a gauge-invariant information of the two-body interaction.

The advantage of working with gauge-invariant quantities is the possibility of comparing results from different approaches.
In fact, besides PN (see Ref. \cite{Blanchet:2013haa} for a review),  there exist other approximation methods which are largely used in the literature, ranging from post-Minkowskian (PM) \cite{Bel:1981be,Damour:2016gwp,Damour:2017zjx,Damour:2019lcq,Damour:2020tta}, gravitational self-force (GSF) (see, e.g., Refs. \cite{Detweiler:2005kq,Poisson:2011nh,Bini:2015bfb,Barack:2018yvs}), effective field theory (EFT) \cite{Levi:2018nxp,Foffa:2013qca,Foffa:2021pkg,Kalin:2020mvi,Kalin:2020fhe,Dlapa:2021npj,Dlapa:2021vgp}, amplitudes \cite{Bjerrum-Bohr:2018xdl,Kosower:2018adc,Cheung:2018wkq,Bjerrum-Bohr:2019kec,Bern:2019nnu,Bern:2021dqo,Bern:2021yeh,Herrmann:2021tct,Manohar:2022dea}, numerical relativity (see, e.g., Ref. \cite{Bishop:2016lgv}). 
A recent approach, termed \lq\lq Tutti Frutti," \cite{Bini:2019nra,Bini:2020wpo,Bini:2020nsb,Bini:2020hmy,Bini:2020rzn} has shown how a combined use of most of the various methods listed above can be the key feature for obtaining (or cross checking) new results. 

The long-term effect of the loss of energy and angular momentum on the binary orbits is to circularize them, as first shown by Peters and Mathews \cite{Peters:1963ux,Peters:1964zz} at the Newtonian level. This is the reason why inspiralling compact binaries are usually modeled as moving in quasi-circular orbits, and much effort has been devoted in the last years to improve the PN amplitude accuracy of waveforms in that situation (see, e.g., Ref. \cite{Faye:2014fra} and references therein for the completion of the full waveform up to the 3.5PN order).
The effect of the eccentricity on the Newtonian orbital-averaged losses is to enhance their quasi-circular values in such a way that they can be written as the product of the radiated energy and angular momentum for a quasi-circular orbit with the same semi-major axis times an \lq\lq enhancement factor,'' which is a function of the eccentricity only, and reduces to unity for vanishing eccentricity.  
Peters-Mathews notion of enhancement factor has been extended to PN corrections in Refs. \cite{Arun:2007rg,Arun:2007sg,Arun:2009mc}, where both energy and angular momentum fluxes were computed at the 3PN order for ellipticlike orbits.
At that level of accuracy the fluxes receive contributions from both instantaneous (i.e., local-in-time) terms and hereditary terms (including tail, tail-of-tail, tail-squared, and memory terms).
The latter arise from the nonlinear interactions between multipoles starting at fractional 1.5PN order, and are expressed as non-local integrals over the full past history of the source \cite{Blanchet:1987wq,Blanchet:1992br,Blanchet:1993ec,Blanchet:1997jj}.
Enhancement factors can be defined also for the hereditary part of the fluxes, but in that case they can be determined only numerically or at least as power series expansions in the small-eccentricity.

The aim of the present work is to extend the notion of enhancement functions to the case of hyperboliclike motion.
The complete (instantaneous plus hereditary) energy flux has been computed very recently at the 3PN order \cite{Cho:2021fmv,Cho:2022pqy}, whereas the angular momentum flux is known only partially at the same approximation level.
In fact, the 2PN instantaneous part and the leading order (Newtonian) tails can be found in Ref. \cite{Bini:2021gat}, whereas higher-order (Newtonian) tails have been computed in Ref. \cite{Bini:2021qvf}. 
Therefore, in order to get the complete 3PN angular momentum flux one would need the instantaneous contribution at 3PN order, and the 1PN corrections to the leading order tails, which are absolute 2.5PN order.
We evaluate here the latter contribution, and use for our analysis both energy and angular momentum fluxes at the same (highest so far) approximation level, i.e., 2.5PN for both. 
The radiated energy and angular momentum will be then parametrized by two different sets of \lq\lq eccentricity enhancement factor functions" (EEFF).

The paper is organized as follows.
In Section II we provide some preliminary information about notation and conventions, including the 2PN-accurate quasi-Keplerian representation of the orbit in harmonic coordinates used to evaluate the orbital average, enough for the purposes of the present work.
Section III is a brief review of the fractionally 2.5PN-accurate expressions for the complete (instantaneous plus hereditary)  energy and angular momentum fluxes.
The explicit computation of the EEFF is done in Sections IV and V in the case of ellipticlike and hyperboliclike motion, respectively.
Instantaneous contributions are straightforwardly computed by working directly in the time domain, whereas tail integrals are more conveniently evaluated in the frequency domain.
Furthermore, in the latter case analytical expressions can be only obtained as power series in the small/large eccentricity for ellipticlike/hyperboliclike orbits.
The evolution of orbital elements under 2.5PN gravitational radiation reaction is discussed in Section VI.
The concluding Section VII summarizes our results, most of which are listed in the Appendices.

\section{Notation and basic definitions}

Let us consider a nonspinning two-body system with masses $m_1$ and $m_2$ (total mass $M=m_1+m_2$, reduced mass $\mu=m_1m_2/M$, symmetric mass ratio $\nu=\mu/M$). Here we will follow notation and conventions of Ref. \cite{Arun:2007rg} and consequently we will assume $m_1>m_2$, defining
\bea
X_1&=&\frac{m_1}{M}=\frac12 (1+\sqrt{1-4\nu})\,,\nonumber\\
X_2&=& \frac{m_2}{M}=\frac12 (1-\sqrt{1-4\nu})\,,
\eea
such that $X_1+X_2=1$.

We will use the 2PN-accurate description of the dynamics of the binary system in harmonic coordinates \cite{DD1981a,D1982}, and the corresponding quasi-Keplerian parametrization \cite{Damour:1988mr,Damour:1990jh,SW1993,Memmesheimer:2004cv,Cho:2018upo}.
Polar coordinates $(r,\phi)$ are introduced in the $x$-$y$ plane, which we will assume to be the orbital plane.
The orbital average will be expressed in terms of two independent orbital elements (semi-major axis and eccentricity, for example), then converted to gauge-invariant functions of the total center-of-mass energy $E$ and angular momentum  $J$ of the binary system, or their dimensionless versions
\beq 
\bar E \equiv \frac{E-Mc^2}{\mu c^2}\,, \qquad
j\equiv \frac{c J}{G M \mu}\,.
\eeq
The total Arnowitt-Deser-Misner (ADM) mass of the system is then given by $\mathcal{M}=Mc^2(1+\nu\bar E)$.
Following the effective-one-body prescriptions we also write
\beq
E=Mc^2 h \,,\quad h=\sqrt{1+2\nu(\gamma-1)}\,,
\eeq
with $\gamma$ the effective energy (per unit of $\mu c^2$) of the system, whence 
\beq
\bar E= \frac{h-1}{\nu}\,.
\eeq
We will find convenient to work with a dimensionless radial distance $r= c^2 r^{\rm phys}/(GM)$ and 
dimensionless rescaled orbital parameters, such as a dimensionless semi-major axis $a_r \equiv c^2 a^{\rm phys}/( GM)$ and a dimensionless radial period $T_r=c^3 T_r^{\rm phys}/(G M)$ for elliptic motion.

The total energy and angular momentum fluxes $\mathcal{F}\equiv\left(\frac{d E}{dU}\right)^{\rm GW}$ and $\mathcal{G}_i\equiv\left(\frac{d J_i}{dU}\right)^{\rm GW}$ can be written as multipolar series parametrized by the radiative multipole moments $U_L$ and $V_L$ evaluated at the retarded time $U$ as follows~\cite{Thorne:1980ru}
\bea
\mathcal{F}&=&\sum_{l=2}^{\infty}\frac{G}{c^{2l+1}}\left[a_lU_L^{(1)}U_L^{(1)}
+\frac1{c^2}b_lV_L^{(1)}V_L^{(1)}\right]
\,,\nonumber\\
\mathcal{G}_i&=&\epsilon_{ijk}\sum_{l=2}^{\infty}\frac{Gl}{c^{2l+1}}\left[a_lU_{jL-1}U_{kL-1}^{(1)}
+\frac1{c^2}b_lV_{jL-1}V_{kL-1}^{(1)}\right]
\,,\nonumber\\
\eea
with coefficients
\beq
a_l=\frac{(l + 1)(l + 2)}{(l - 1)ll!(2l + 1)!!}\,,\qquad
b_l=a_l\left(\frac{2l}{l+1} \right)^2\,,
\eeq
where $L=i_1i_2\cdots i_l$ is a multi-index, and a superscript in parenthesis denotes repeated retarded time derivatives. 
These radiative moments are then related to the intermediate canonical moments ($M_L, S_L$), which are useful to describe the nonlinear interaction between multipole moments \cite{Blanchet:1997jj,Blanchet:2008je}
\beq
U_L=M_L^{(l)}+ O \left(\frac{G}{c^3}\right)
\,, \qquad
V_L=S_L^{(l)}+ O \left(\frac{G}{c^3}\right)\,.
\eeq
Such a construction leads to two different types of terms: instantaneous terms, in which all the canonical moments are evaluated at the current (retarded) time, and hereditary terms, which are given by nonlocal integrals extending over the past history of the source, and comprise tail, tail-of-tail, tail squared, and nonlinear memory terms.
For instance, the mass moment reads
\beq
\label{decUL}
U_L=U_L^{\rm inst}+U_L^{\rm hered}\,,
\eeq
with
\beq
\label{decULh}
U_L^{\rm hered}=U_L^{\rm tail}+U_L^{\rm tail(tail)}+U_L^{\rm (tail)^2}+U_L^{\rm mem}+\ldots\,,
\eeq
up to the cubic order in $G$, and higher nonlinear interaction terms (with increasing powers of $G$).
Finally, the canonical moments are in turn expressed in terms of the source moments ($I_L, J_L$) evaluated at the retarded time $u=t-r/c$ in harmonic coordinates, which is related to the retarded time $U$ in radiative coordinates by 
\beq
\label{Uvst}
U = u
-\frac{2G\mathcal{M}}{c^3}\ln\left(\frac{r}{r_0}\right) +  O \left(\frac{G^2}{c^5}\right)\,,
\eeq
where $r_0$ denotes a constant length scale.
The difference between these two sets of multipoles is 2.5PN order, i.e.,
\beq
M_L=I_L+ O \left(\frac{G}{c^5}\right)
\,, \qquad
S_L=J_L+ O \left(\frac{G}{c^5}\right)\,.
\eeq
Therefore, the final expressions relating the radiative moments with the source moments involve a double PM-PN expansion.
The same decomposition \eqref{decUL}--\eqref{decULh} holds for the energy and angular momentum fluxes.

In general, throughout the paper we will set $G=M=c=1$ to ease notation.
However, we will use the placeholder $\eta \equiv \frac1c$ to keep track of the fractional PN order of the various contributions to PN-expanded quantities ($\eta^0$ stands for the Newtonian level of accuracy,  $\eta^2$ for the 1PN, etc.).

\subsection{Ellipticlike orbits}

The 2PN-accurate quasi-Keplerian parametrization of the ellipticlike motion is 
\begin{eqnarray} \label{QK}
r&=& a_r (1-e_r \cos u)\,,\nonumber\\
\ell&=&nt= u-e_t \sin u + f_t \sin V+g_t(V-u)\,,\nonumber\\
\phi &=&K[V+f_\phi \sin 2V+g_\phi \sin 3V]\,,
\end{eqnarray}
with
\beq
V(u)=2 \,{\rm arctan}\left[\sqrt{\frac{1+e_\phi}{1-e_\phi}}\tan \frac{u}{2}  \right]\,.
\eeq
Here $a_r$ is the semi-major axis of the orbit, $e_t,e_r,e_\phi$ are three  kinds of eccentricities, $K=1+k$ is the periastron advance and $n = \frac{2\pi}{T_r}$ is the frequency of the radial motion. Their expressions in harmonic coordinates are listed, e.g., in Table II of Ref. \cite{Bini:2020nsb} as functions of the specific binding energy $\bar E$ and of the dimensionless  angular momentum $j$ of the system. 

The orbital average of a generic function $F(t)$ over a period of radial motion is given by
\bea
\langle  F \rangle_{\rm (ell)}&=&\frac{1}{T_r}\int_0^{T_r} dt F(t)\nonumber\\ 
&=& \frac{1}{2\pi}\int_0^{2\pi} d\ell F(\ell)\,.
\eea

The motion is in general doubly-periodic, since the angle $\ell$ measures the periodicity in the radial motion, whereas the angle $g=k\ell$ measures the mean periastron precession.
The azimuthal motion can then be expressed as $\phi(l,g)=\ell+g+W(\ell)$, with $W(\ell)$ a periodic function of $\ell$.
Hence, the various multipole moments still admit a decomposition as Fourier series (see Eqs. (3.17a) and (3.17b) of Ref. \cite{Arun:2007rg}).
The usual single-periodic Fourier decomposition 
\bea
\label{Nseriesellip}
X_L(t)&=&\sum_{p=-\infty}^\infty e^{ip\ell}\hat X_L(p)\,,\nonumber\\ 
{}\hat X_L(p)&=&\frac1{2\pi}\int_0^{2\pi} d\ell e^{-ip\ell}X_L(\ell)\,,
\eea
is recovered at the Newtonian approximation, where $k\to0$.

\subsection{Hyperboliclike orbits}

The 2PN-accurate quasi-Keplerian parametrization of the hyperboliclike motion is 
\begin{eqnarray} \label{hypQK2PN}
r&=& \bar a_r (e_r \cosh v-1)\,,\nonumber\\
\bar n t&=&e_t \sinh v-v + f_t V+g_t \sin V\,,\nonumber\\
\phi &=&K[V+f_\phi \sin 2V+g_\phi \sin 3V]\,,
\end{eqnarray}
with
\beq
\label{Vdef}
V(v)=2\, {\rm arctan}\left[\sqrt{\frac{e_\phi+1}{e_\phi-1}}\tanh \frac{v}{2}  \right]\,.
\eeq
The expressions of the orbital parameters $\bar n$, $\bar a_r$, $K$, $e_t,e_r,e_\phi$, $f_t,g_t,f_\phi, g_\phi$ are given, e.g., in Table VIII of Ref. \cite{Bini:2020hmy} in harmonic coordinates as functions of the conserved energy and angular momentum of the system.

In place of $\bar E$ we will often use the linear momentum at infinity $p_{\infty}$, such that
\bea
\label{barE_pinf}
\bar E&=&\frac{1}{\nu }[\sqrt{1+2\nu (\gamma -1)}-1] \nonumber\\
&\approx & \frac12 p_{\infty}^2 - \frac18 (1+ \nu) p_{\infty}^4  \nonumber\\
&&+\frac{1}{16} (1+ \nu+  \nu^2)p_{\infty}^6 +O(p_{\infty}^8)\,,
\eea
with $\gamma=\sqrt{1+p_{\infty}^2}$, and inverse relation
\bea
p_{\infty}^2&=&\frac14   \bar E (\bar E\nu + 2) (\bar E^2\nu + 2\bar E + 4) \nonumber\\
&\approx & 2\bar E + (1+\nu)\bar E^2  + \nu\bar E^3 +O(\bar E^4)\,.
\eea

The orbital average of a generic function $F(t)$ over the full scattering process is given by
\beq
\langle F \rangle_{\rm(hyp)}= \int_{-\infty}^{\infty}dt\, F (t)\,.
\eeq
The various multipole moments can be expanded as Fourier integrals
\bea
X_L(t)&=&\int_{-\infty}^\infty \frac{d\omega}{2\pi} e^{-i\omega t}\hat X_L(\omega)\,,\nonumber\\
{}\hat X_L(\omega)&=&\int_{-\infty}^\infty dt e^{i\omega t}X_L(t)\,.
\eea

\section{2.5PN accurate energy and angular momentum fluxes}

The fractionally 2.5PN-accurate expressions for the instantaneous part of the energy and angular momentum fluxes read
\begin{eqnarray}
\label{flux2PNdef}
{\mathcal F}^{\rm inst}&=& \frac{G}{c^5} \left[\frac15 I_{ab}^{\rm (3)} I_{ab}^{\rm (3)}\right.\nonumber\\
&+&\eta^2\left(\frac1{189 } I_{abc}^{\rm (4)} I_{abc}^{\rm (4)} +\frac{16}{45 } J_{ab}^{\rm (3)} J_{ab}^{\rm (3)}\right)\nonumber\\
&+&\left.
\eta^4\left(\frac{1}{9072}I_{abcd}^{\rm (5)} I_{abcd}^{\rm (5)} +\frac{1}{84}J_{abc}^{\rm (4)} J_{abc}^{\rm (4)}\right)
\right]
+{\mathcal F}^{\rm inst}_{\rm 2.5PN}
\,,\nonumber\\
{\mathcal G}_i^{\rm inst}&=& \frac{G}{c^5}\epsilon_{iab} \left[\frac25 I_{aj}^{\rm (2)} I_{bj}^{\rm (3)}\right.\nonumber\\
&+&\eta^2\left(\frac1{63} I_{ajk}^{\rm (3)} I_{bjk}^{\rm (4)} +\frac{32}{45 } J_{aj}^{\rm (2)} J_{bj}^{\rm (3)}\right)\nonumber\\
&+&\left.
\eta^4\left(\frac{1}{2268}I_{ajkl}^{\rm (4)} I_{bjkl}^{\rm (5)} +\frac{1}{28}J_{ajk}^{\rm (3)} J_{bjk}^{\rm (4)}\right)
\right]
+{\mathcal G}_{i\,\rm 2.5PN}^{\rm inst}
\,,
\nonumber\\
\end{eqnarray}
respectively, where the 2.5PN terms are given in Eqs. (5.2d) of Ref. \cite{Arun:2007rg} (energy) and (2.7) of Ref. \cite{Arun:2009mc} (angular momentum).
The orbital average of the latter terms is vanishing along both ellipticlike and hyperboliclike orbits, $\langle {\mathcal F}^{\rm inst}_{\rm 2.5PN}\rangle=0=\langle {\mathcal G}_{i\,\rm 2.5PN}^{\rm inst}\rangle$ (as explicitly shown in Ref. \cite{Arun:2007rg} (see text below Eq. (8.3) there) in the former case, but straightforwardly extended to the hyperbolic case too).

The tail part starting at the fractional 1.5PN order reads 
\bea
\label{F_G_tail}
{\mathcal F}^{\rm tail}&=& \frac{G^2{\mathcal M}}{c^8}\left\{\frac45 I_{ij}^{(3)}(t)\int_0^\infty d\tau I_{ij}^{(5)}(t-\tau)\ln \left( \frac{\tau}{C_{I_2}} \right)\right.\nonumber\\
&+&\frac{1}{c^2}\left[
\frac{4}{189 }I_{ijk}^{(4)}(t)
\int_0^\infty d\tau I_{ijk}^{(6)}(t-\tau)\ln \left( \frac{\tau}{C_{I_3}} \right)\right.\nonumber\\
&+&\left.\left.
\frac{64 }{45}J_{ij}^{(3)}(t)
\int_0^\infty d\tau J_{ij}^{(5)}(t-\tau)\ln \left( \frac{\tau}{C_{J_2}} \right)
\right]\right\}
\,,\nonumber\\
{\mathcal G}_{i}^{\rm tail}&=& \frac{G^2{\mathcal M}}{c^8}\epsilon_{iab}
\left\{\frac45 I_{aj}^{(2)}(t)\int_0^\infty d\tau I_{ij}^{(5)}(t-\tau)\ln \left( \frac{\tau}{C_{I_2}} \right)\right.\nonumber\\
&+&
\frac45 I_{bj}^{(3)}(t)\int_0^\infty d\tau I_{aj}^{(4)}(t-\tau)\ln \left( \frac{\tau}{C_{I_2}} \right)
\nonumber\\
&+&\frac{1}{c^2}\left[
\frac{64}{45}J_{aj}^{(2)}(t)
\int_0^\infty d\tau J_{bj}^{(5)}(t-\tau)\ln \left( \frac{\tau}{C_{J_2}} \right)\right.\nonumber\\
&+&
\frac{64 }{45}J_{bj}^{(3)}(t)
\int_0^\infty d\tau J_{aj}^{(4)}(t-\tau)\ln \left( \frac{\tau}{C_{J_2}} \right)\nonumber\\
&+&
\frac{2}{63}I_{ajk}^{(3)}(t)
\int_0^\infty d\tau I_{bjk}^{(6)}(t-\tau)\ln \left( \frac{\tau}{C_{I_3}} \right)\nonumber\\
&+&\left.\left.
\frac{2}{63}I_{bjk}^{(4)}(t)
\int_0^\infty d\tau I_{ajk}^{(5)}(t-\tau)\ln \left( \frac{\tau}{C_{I_3}} \right)
\right]\right\}
\,,
\eea
as in Eq. 2.9 of Ref. \cite{Arun:2007rg} and Eq. 2.9 of Ref. \cite{Arun:2009mc}, respectively.
Here we have introduced the following set of multipolar constants ($\tau_0=cr_0$)
\bea
C_{I_2}&=& 2\tau_0 e^{-11/12}
\,,\nonumber\\
C_{I_3}&=& 2\tau_0 e^{-97/60}
\,,\nonumber\\
C_{J_2}&=& 2\tau_0 e^{-7/6}
\,.
\eea

At the fractional 2.5PN order there is a further contribution to the angular momentum flux coming from the following nonlinear memory integral
\beq
\label{Gimemory}
{\mathcal G}_i^{\rm mem}=
\frac{4}{35}\frac{G^2}{c^{10}}\,\epsilon_{iab}\,I_{aj}^{{(3)}}(t)\int_{-\infty}^{t} d \tau I_{c b}^{{(3)}}(\tau)\, I_{j c}^{{(3)}}(\tau)
\,.
\eeq
Orbital averaging of ${\cal G}_i^{\rm mem}$ gives a vanishing contribution in the bound case \cite{Arun:2007rg}, whereas in the unbound case it has been computed in Ref. \cite{Bini:2021qvf}.

To summarize, at the 2.5PN level of accuracy the radiated energy and angular momentum read
\beq
\Delta E\equiv\langle {\mathcal F}\rangle=\langle {\mathcal F}^{\rm inst}\rangle+\langle {\mathcal F}^{\rm hered}\rangle
\,,
\eeq
with
\beq
\langle {\mathcal F}^{\rm hered}\rangle=\langle {\mathcal F}^{\rm tail}\rangle\,,
\eeq
and
\beq
\Delta J_i\equiv\langle {\mathcal G}_i\rangle=\langle {\mathcal G}_i^{\rm inst}\rangle+\langle {\mathcal G}_i^{\rm hered}\rangle
\,,
\eeq
with
\beq
\langle {\mathcal G}_i^{\rm hered}\rangle=\langle {\mathcal G}_i^{\rm tail}\rangle+\langle {\mathcal G}_i^{\rm mem}\rangle
\,,
\eeq
respectively.

\section{Multipolar invariants along ellipticlike orbits}

\subsection{Instantaneous part}

The orbital average of the instantaneous fluxes \eqref{flux2PNdef} can be written as
\begin{eqnarray}
\label{flux2PNell}
(\Delta E)_{\rm inst}^{\rm (ell)}&=& \frac{G}{c^5}\frac{n}{2\pi} \left[\frac15 I^{\rm (ell)}_{2[33]}\right.\nonumber\\
&+&
\eta^2\left(\frac1{189 } I^{\rm (ell)}_{3[44]} +\frac{16}{45 } J^{\rm (ell)}_{2[33]}\right)\nonumber\\
&+&\left.
\eta^4\left(\frac{1}{9072}I^{\rm (ell)}_{4[55]} +\frac{1}{84}J^{\rm (ell)}_{3[44]}\right)
\right]
\,,\nonumber\\
(\Delta J_i)_{\rm inst}^{\rm (ell)}&=& \frac{G}{c^5}\frac{n}{2\pi}\left[\frac25 I^{*\,\rm (ell)}_{i\,2[2,3]}\right.\nonumber\\
&+&
\eta^2\left(\frac1{63} I^{*\,\rm (ell)}_{i\,3[3,4]} +\frac{32}{45 } J^{*\,\rm (ell)}_{i\,2[2,3]}\right)\nonumber\\
&+&\left.
\eta^4\left(\frac{1}{2268}I^{*\,\rm (ell)}_{i\,4[4,5]} +\frac{1}{28}J^{*\,\rm (ell)}_{i\,3[3,4]}\right)
\right]
\,,
\end{eqnarray}
where we have introduced the notation
\bea
\label{notation_ellip}
X^{\rm (ell)}_{l[r,s]}&=&\int_0^{T_r} dt X_L^{(r)}(t)X_L^{(s)}(t)
\,,\nonumber\\
X^{*\,\rm (ell)}_{i\,l[r,s]}&=&\epsilon_{iab}\int_0^{T_r} dt X_{aL-1}^{(r)}(t)X_{bL-1}^{(s)}(t)
\,.
\eea
These integrals are more easily computed in the time domain than in the frequency domain.

\subsubsection{Energy EEFF}

Let us consider for example the first term in the energy flux
\beq
I^{\rm (ell)}_{2[3,3]}=  \int_0^{2\pi} \frac{dt}{du}I_{ij}^{(3)}(t(u))I_{ij}^{(3)}(t(u)) du \,,
\eeq
turns out to be
\beq
\label{I2_33}
I^{\rm (ell)}_{2[3,3]}= 
64\frac{\pi\nu^2}{a_r^{7/2}}\left[f(e_t) +\frac{\eta^2}{a_r}f_{\rm 1PN}(e_t;\nu)
+\frac{\eta^4}{a_r^2}f_{\rm 2PN}(e_t;\nu)\right]\,,
\eeq
at the 2PN level of accuracy.
Here $f(e_t)$ is the Peters-Mathews EEFF
\beq
f(e_t)=\frac{1+ \frac{73}{24}e_t^2+\frac{37}{96}e_t^4}{(1-e_t^2)^{7/2}}\,,
\eeq
with PN corrections $f_{\rm 1PN}(e_t;\nu)$ and $f_{\rm 2PN}(e_t;\nu)$ given by
\begin{widetext}
\bea
f_{\rm 1PN}(e_t;\nu)&=&\frac{ 
-\frac{235}{42}+\frac{53\nu }{14}
+\left(\frac{5081}{336}+\frac{1135 \nu }{112}\right)e_t^2
+\left(\frac{3291}{64}-\frac{199 \nu }{64}\right)e_t^4
+\left(\frac{1657}{336}-\frac{809 \nu }{1344}\right)e_t^6
}{(1-e_t^2)^{9/2}}
\,,\nonumber\\
f_{\rm 2PN}(e_t;\nu)&=& 
\frac{1}{(1-e_t^2)^{11/2}}\left\{
\frac{101765 \nu^2}{10584}-\frac{203921 \nu }{10584}+\frac{202513}{10584}
+\left(\frac{153761 \nu ^2}{12096}+\frac{462463 \nu}{12096}-\frac{1870559}{12096}\right)e_t^2\right.\nonumber\\
&&
+\left(-\frac{5902535 \nu ^2}{112896}+\frac{5622539 \nu}{112896}+\frac{21233029}{112896}\right)e_t^4
+\left(-\frac{8647 \nu ^2}{896}-\frac{49255 \nu}{448}+\frac{647945}{1152}\right)e_t^6\nonumber\\
&&
+\left(\frac{16007 \nu ^2}{37632}-\frac{239207 \nu}{25088}+\frac{6287893}{150528}\right)e_t^8\nonumber\\
&&\left.
+e_t^2(1-e_t^2)^{1/2}\left[\frac{785}{8}-\frac{157 \nu }{4} 
+ \left(\frac{1005}{8}-\frac{201 \nu }{4}\right)e_t^2 
+ \left(\frac{555}{64}-\frac{111 \nu }{32}\right)e_t^4\right]
\right\}
\,,
\eea
respectively.
$f_{\rm 1PN}(e_t;\nu)$ is a linear function of $\nu$, whereas $f_{\rm 2PN}(e_t;\nu)$ is quadratic.
Therefore, following Ref. \cite{Arun:2007rg} one can introduce a EEFF $f_{\rm nPN}^{\nu^k}(e_t)$ for every order in $\nu$.
Eq. \eqref{I2_33} can then be rewritten as
\bea
I^{\rm (ell)}_{2[3,3]}&=& 
64\frac{\pi\nu^2}{a_r^{7/2}}\left[f(e_t) +\frac{\eta^2}{a_r}\left(-\frac{235}{42}f_{\rm 1PN}^{\nu^0}(e_t) +\frac{53}{14}\nu f_{\rm 1PN}^{\nu^1}(e_t)\right)\right.\nonumber\\
&+& \left. 
\frac{\eta^4}{a_r^2}\left(\frac{202513}{10584}f_{\rm 2PN}^{\nu^0}(e_t)-\frac{203921}{10584}\nu f_{\rm 2PN}^{\nu^1}(e_t)+\frac{101765}{10584}\nu^2 f_{\rm 2PN}^{\nu^2}(e_t)\right)\right]\,.
\eea

We list below the expressions for the remaining terms entering the orbital average of the instantaneous energy flux \eqref{flux2PNell} up to the needed PN accuracy
\bea
I^{\rm (ell)}_{3[4,4]}&=& 
\frac{16404}{5}\frac{\pi\nu^2}{a_r^{9/2}}(1-4\nu)\left[F(e_t) +\frac{\eta^2}{a_r}F_{\rm 1PN}(e_t;\nu)
+O(\eta^4)\right]
\,,\nonumber\\
J^{\rm (ell)}_{2[3,3]}&=& 
\frac{\pi\nu^2}{a_r^{9/2}}(1-4\nu)\left[{\mathcal F}(e_t) +\frac{\eta^2}{a_r}{\mathcal F}_{\rm 1PN}(e_t;\nu)
+O(\eta^4)\right]
\,,\nonumber\\
I^{\rm (ell)}_{4[5,5]}&=& 
\frac{1836032}{7}\frac{\pi\nu^2}{a_r^{11/2}}(1-3\nu)^2\left[{\mathfrak F}(e_t) 
+O(\eta^2)\right]
\,,\nonumber\\
J^{\rm (ell)}_{3[4,4]}&=& 
\frac{256}{3}\frac{\pi\nu^2}{a_r^{11/2}}(1-3\nu)^2\left[{\mathbb F}(e_t) 
+O(\eta^2)\right]
\,,
\eea
where
\bea
F(e_t)&=& 
\frac{1 + \frac{18509}{2734} e_t^2 + \frac{50295}{10936} e_t^4 + \frac{5091}{21872} e_t^6}{(1-e_t^2)^{9/2}}
\,,\nonumber\\
{\mathcal F}(e_t)&=&
\frac{1 + \frac{19}{2} e_t^2 + \frac{69}{8} e_t^4 +\frac{9}{16} e_t^6}{(1-e_t^2)^{9/2}} 
\,,\nonumber\\
{\mathfrak F}(e_t)&=& 
\frac{1+ \frac{3377059}{286880} e_t^2+ \frac{21045417}{1147520} e_t^4+ \frac{11528181}{2295040} e_t^6 + \frac{2347659}{18360320} e_t^8}{(1-e_t^2)^{11/2}}
\,,\nonumber\\
{\mathbb F}(e_t)&=&
\frac{1+ \frac{2267}{160} e_t^2+ \frac{17249}{640} e_t^4+ \frac{11693}{1280} e_t^6 + \frac{2979}{10240} e_t^8}{(1-e_t^2)^{11/2}}
\,,
\eea
and PN corrections
\bea
F_{\rm 1PN}(e_t;\nu)&=&\frac1{(1-e_t^2)^{11/2}}\left[
\frac{45107 \nu}{8202}-\frac{77917}{8202}
+\left(\frac{575915 \nu }{16404}-\frac{439297}{16404}\right)e_t^2
+\left(\frac{234293 \nu }{21872}+\frac{2247529}{21872}\right)e_t^4\right.\nonumber\\
&&\left.
+\left(\frac{3149207}{43744}-\frac{349773 \nu }{43744}\right)e_t^6
+\left(\frac{269637}{87488}-\frac{18807 \nu }{43744}\right)e_t^8
\right]
\,,\nonumber\\
{\mathcal F}_{\rm 1PN}(e_t;\nu)&=&\frac1{(1-e_t^2)^{11/2}}\left[
\frac{41 \nu}{14}-\frac{19}{7}
+\left(\frac{285 \nu }{28}-\frac{139}{28}\right)e_t^2
+\left(\frac{13299}{112}-\frac{5395 \nu }{112}\right)e_t^4\right.\nonumber\\
&&\left.
+\left(\frac{22915}{224}-\frac{8877 \nu }{224}\right)e_t^6
+\left(\frac{10251}{1792}-\frac{873 \nu }{448}\right)e_t^8
\right]
\,.
\eea

\end{widetext}

The above expressions for the Newtonian EEFF can be also obtained by working in the Fourier domain. 
Inserting the decomposition \eqref{Nseriesellip} into Eq. \eqref{notation_ellip} yields

\bea
\frac{n}{2\pi}X^{\rm (ell)}_{l[r,s]}&=&  \int_0^{2\pi} \frac{d\ell}{2\pi} \sum_{p=-\infty}^\infty (ipn)^r e^{ip\ell}\hat X_L(p)
\nonumber\\
&&\times
\sum_{q=-\infty}^\infty (iqn)^s e^{iq\ell}\hat X_L(q)\nonumber\\
&=& i^{r+s} n^{r+s} \left[(-1)^s  +(-1)^r \right]S_{r+s,X_L}^{\rm (ell)}
\,,\nonumber\\
\eea
where we have denoted
\bea
\label{S_no_log}
S_{n,X_L}^{\rm (ell)}&=&\sum_{p=1}^\infty p^n \hat X_L(p)\hat X_L(-p)\,.
\eea
One should distinguish the cases $r,s$=even, odd; for example, the above expression vanishes when $r$=even and $s$=odd (and  $r$=odd and $s$=even), while when both $r$ and $s$ are even becomes
\beq
\label{time_dom}
\frac{n}{2\pi}X^{\rm (ell)}_{l[r,s]}
=(-n)^{r+s} S_{r+s,X_L}^{\rm (ell)}\,,\qquad r,s={\rm even}.
\eeq
In the special case
$r=s$ (diagonal case), we find
\beq
\frac{n}{2\pi} X^{\rm (ell)}_{l[r,r]}
=2 n^{2r}   S_{2r,X_L}^{\rm (ell)}\,,
\eeq
so that
\bea
f(e_t)&=&  \frac{1}{16\mu^2a_r^4}S_{6,I_2}^{\rm (ell)}
\,,\nonumber\\
F(e_t)&=& \frac{5}{4101(1-4\nu)\mu^2a_r^6}S_{8,I_3}^{\rm (ell)} 
\,,\nonumber\\
{\mathcal F}(e_t)&=& \frac{4}{(1-4\nu)\mu^2a_r^3}S_{6,J_2}^{\rm (ell)} 
\,,\nonumber\\
{\mathfrak F}(e_t)&=& \frac{7}{459008(1-3\nu)^2\mu^2a_r^8}S_{10,I_4}^{\rm (ell)} 
\,,\nonumber\\
{\mathbb F}(e_t)&=& \frac{3}{64(1-3\nu)^2\mu^2a_r^5}S_{8,J_3}^{\rm (ell)}
\,. 
\eea

\subsubsection{Angular momentum EEFF}

The multipolar invariants entering the angular momentum flux \eqref{flux2PNell} can be parametrized in terms of EEFF just as in the case of the radiated energy, labelled by a star in the equations below
\bea
I^{*\,\rm (ell)}_{z\,2[2,3]}&=& 
32\frac{\pi\nu^2}{a_r^2}\left[f^{*}(e_t) +\frac{\eta^2}{a_r}f^{*}_{\rm 1PN}(e_t;\nu)\right.\nonumber\\
&&\left.
+\frac{\eta^4}{a_r^2}f^{*}_{\rm 2PN}(e_t;\nu)\right]
\,,\nonumber\\
I^{*\,\rm (ell)}_{z\,3[3,4]}&=& 
\frac{10936}{10}\frac{\pi\nu^2}{a_r^3}(1-4\nu)\left[F^{*}(e_t) +\frac{\eta^2}{a_r}F^{*}_{\rm 1PN}(e_t;\nu)\right.\nonumber\\
&&\left.
+O(\eta^4)\right]
\,,\nonumber\\
J^{*\,\rm (ell)}_{z\,2[2,3]}&=& 
\frac12\frac{\pi\nu^2}{a_r^3}(1-4\nu)\left[{\mathcal F}^{*}(e_t) +\frac{\eta^2}{a_r}{\mathcal F}^{*}_{\rm 1PN}(e_t;\nu)\right.\nonumber\\
&&\left.
+O(\eta^4)\right]
\,,\nonumber\\
I^{*\,\rm (ell)}_{z\,4[4,5]}&=& 
\frac{459008}{7}\frac{\pi\nu^2}{a_r^4}(1-3\nu)^2\left[{\mathfrak F}^{*}(e_t) 
+O(\eta^2)\right]
\,,\nonumber\\
J^{*\,\rm (ell)}_{z\,3[3,4]}&=& 
\frac{256}{9}\frac{\pi\nu^2}{a_r^4}(1-3\nu)^2\left[{\mathbb F}^{*}(e_t) 
+O(\eta^2)\right]
\,,
\eea
where
\bea
f^{*}(e_t)&=&
\frac{1+ \frac{7}{8}e_t^2}{(1-e_t^2)^2}
\,,\nonumber\\
F^{*}(e_t)&=& 
\frac{1 + \frac{4161}{1367} e_t^2 + \frac{5301}{10936} e_t^4}{(1-e_t^2)^3}
\,,\nonumber\\
{\mathcal F}^{*}(e_t)&=&
\frac{1 + 3 e_t^2 + \frac{3}{8} e_t^4}{(1-e_t^2)^3} 
\,,\nonumber\\
{\mathfrak F}^{*}(e_t)&=& 
\frac{1+ \frac{372981}{57376}e_t^2+  \frac{490935}{114752}e_t^4+  \frac{107829}{459008}e_t^6 }{(1-e_t^2)^4}
\,,\nonumber\\
{\mathbb F}^{*}(e_t)&=&
\frac{1+ \frac{229}{32} e_t^2+ \frac{327}{64} e_t^4+ \frac{69}{256} e_t^6}{(1-e_t^2)^4}
\,,
\eea
with PN corrections
\begin{widetext}
\bea
f^{*}_{\rm 1PN}(e_t;\nu)&=&\frac{ 
-\frac{86}{21}+\frac{23 \nu }{7}
+\left(\frac{1957}{168}+\frac{157 \nu }{56}\right)e_t^2
+\left(\frac{5179}{672}-\frac{87 \nu }{112}\right)e_t^4
}{(1-e_t^2)^3}
\,,\nonumber\\
f^{*}_{\rm 2PN}(e_t;\nu)&=&\frac{1}{(1-e_t^2)^4}\left\{
\frac{10051 \nu^2}{1323}-\frac{96133 \nu }{5292}+\frac{29413}{2646}
+\left(-\frac{443 \nu ^2}{252}+\frac{41113 \nu }{2016}-\frac{168043}{2016}\right)e_t^2\right.\nonumber\\
&&
+\left(-\frac{5767 \nu ^2}{441}+\frac{28331 \nu }{14112}+\frac{1428527}{14112}\right)e_t^4
+\left(\frac{407 \nu ^2}{1568}-\frac{88253 \nu }{9408}+\frac{1830295}{37632}\right)e_t^6\nonumber\\
&&\left.
+e_t^2(1-e_t^2)^{1/2}\left[\frac{345}{8}-\frac{69 \nu }{4}
+ \left(\frac{105}{8}-\frac{21 \nu }{4}\right)e_t^2\right]
\right\}
\,,\nonumber\\
F^{*}_{\rm 1PN}(e_t;\nu)&=&\frac1{(1-e_t^2)^4}\left[
\frac{20503 \nu }{4101}-\frac{32807}{4101}
+\left(\frac{20874\nu }{1367}-\frac{6684}{1367}\right)e_t^2
+\left(\frac{354805}{10936}-\frac{6293 \nu }{10936}\right)e_t^4\right.\nonumber\\
&&\left.
+\left(\frac{49539}{10936}-\frac{6657 \nu }{10936}\right)e_t^6
\right]
\,,\nonumber\\
{\mathcal F}^{*}_{\rm 1PN}(e_t;\nu)&=&\frac1{(1-e_t^2)^4}\left[
\frac{17 \nu }{7}-\frac{17}{14}
+\left(\frac{451}{28}-\frac{5 \nu}{7}\right)e_t^2
+\left(\frac{2967}{112}-\frac{573 \nu }{56}\right)e_t^4
+\left(\frac{453}{224}-\frac{6 \nu }{7}\right)e_t^6
\right]
\,.
\eea

\end{widetext}

\subsection{Tail part}

The energy and angular momentum tails (see Eqs.  \eqref{F_G_tail} for their definitions) are known at the 1PN fractional accuracy level \cite{Arun:2007rg,Arun:2009mc}. For example, in Eq. (6.2) of Ref. \cite{Arun:2007rg} one reads the tail contribution to the energy flux written in terms of the EEFF $\varphi(e_t)$, $\psi(e_t)$ and $\zeta(e_t)$.
The latter are known semi-analytically, in the sense that only the first few terms have been given analytically in a small-eccentricity expansion, while most of the information comes from fit of numerical values.

\section{Multipolar invariants along hyperboliclike orbits}

\subsection{Instantaneous part}

The orbital average of the instantaneous fluxes \eqref{flux2PNdef} can be written as
\begin{eqnarray}
\label{flux2PNhyp}
(\Delta E)_{\rm inst}^{\rm (hyp)}&=& \frac{G}{c^5}\left[\frac15 I^{\rm (hyp)}_{2[33]}\right.\nonumber\\
&+&
\eta^2\left(\frac1{189 } I^{\rm (hyp)}_{3[44]} +\frac{16}{45 } J^{\rm (hyp)}_{2[33]}\right)\nonumber\\
&+&\left.
\eta^4\left(\frac{1}{9072}I^{\rm (hyp)}_{4[55]} +\frac{1}{84}J^{\rm (hyp)}_{3[44]}\right)
\right]
\,,\nonumber\\
(\Delta J_i)_{\rm inst}^{\rm (hyp)}&=& \frac{G}{c^5}\left[\frac25 I^{*\,\rm (hyp)}_{i\,2[2,3]}\right.\nonumber\\
&+&
\eta^2\left(\frac1{63} I^{*\,\rm (hyp)}_{i\,3[3,4]} +\frac{32}{45 } J^{*\,\rm (hyp)}_{i\,2[2,3]}\right)\nonumber\\
&+&\left.
\eta^4\left(\frac{1}{2268}I^{*\,\rm (hyp)}_{i\,4[4,5]} +\frac{1}{28}J^{*\,\rm (hyp)}_{i\,3[3,4]}\right)
\right]
\,,
\end{eqnarray}
where we have introduced the notation
\bea
\label{notation_hyp}
X^{\rm (hyp)}_{l[r,s]}&=&\int_{-\infty}^\infty dt X_L^{(r)}(t)X_L^{(s)}(t)
\,,\nonumber\\
X^{*\,\rm (hyp)}_{i\,l[r,s]}&=&\epsilon_{iab}\int_{-\infty}^\infty dt X_{aL-1}^{(r)}(t)X_{bL-1}^{(s)}(t)
\,.
\eea
These integrals are easily computed in the time domain.

\subsubsection{Energy EEFF}

The first term in the energy flux turns out to be
\begin{widetext}
\beq
\label{I2_33hyp}
I^{\rm (hyp)}_{2[3,3]}= 
\frac{\nu^2}{\bar a_r^{7/2}}\left[f^{\rm (hyp)}(e_r) +\frac{\eta^2}{\bar a_r}f^{\rm (hyp)}_{\rm 1PN}(e_r;\nu)
+\frac{\eta^4}{\bar a_r^2}f^{\rm (hyp)}_{\rm 2PN}(e_r;\nu)\right]\,,
\eeq
with
\bea
f^{\rm (hyp)}(e_r)&=& 
\left(
\frac{850}{ 3 (e_r^2 - 1)^{7/2}} 
+ \frac{244}{(e_r^2 - 1)^{5/2}}
+ \frac{74}{ 3 (e_r^2 - 1)^{3/2}}\right) {\rm arccos }\left(-\frac{1}{e_r}\right) 
+ \frac{850}{ 3 (e_r^2 - 1)^3 } 
+ \frac{1346}{ 9 (e_r^2 - 1)^2 }
\,,\nonumber\\
f^{\rm (hyp)}_{\rm 1PN}(e_r;\nu)&=&
\left(
\frac{\frac{1522\nu}{21} + \frac{412}{21}}{(e_r^2 - 1)^{3/2}}
+ \frac{\frac{11385\nu}{7} - \frac{6567}{7}}{ (e_r^2 - 1)^{5/2} }
+ \frac{4940\nu - \frac{12950}{3}}{ (e_r^2 - 1)^{7/2}} 
+ \frac{ \frac{10885\nu}{3} - 3717}{ (e_r^2 - 1)^{9/2}}
\right){\rm arccos }\left(-\frac{1}{e_r}\right)\nonumber\\
&&
+ \frac{\frac{4938\nu}{7} - \frac{76436}{315}}{ (e_r^2 - 1)^2} 
+ \frac{\frac{33575\nu}{9} - \frac{9233}{3}}{ (e_r^2 - 1)^3} 
+ \frac{\frac{10885\nu}{3} - 3717}{ (e_r^2 - 1)^4}
\,,\nonumber\\
f^{\rm (hyp)}_{\rm 2PN}(e_r;\nu)&=&
\left(
\frac{\frac{7073}{147}\nu^2 + \frac{45511}{392}\nu + \frac{164629}{2352}}{ (e_r^2 - 1)^{3/2}} 
+ \frac{\frac{516941}{147}\nu^2 - \frac{1191707}{196}\nu + \frac{1824097}{1764}}{ (e_r^2 - 1)^{5/2}}
\right.\nonumber\\
&&
+ \frac{\frac{42556525}{1764}\nu^2 - \frac{96608923}{1764}\nu + \frac{48504467}{3528}}{ (e_r^2 - 1)^{7/2}} 
+ \frac{\frac{2507479}{54}\nu^2 - \frac{11923121}{108}\nu + \frac{3960697}{108}}{ (e_r^2 - 1)^{9/2}} \nonumber\\
&&\left.
+ \frac{\frac{2217065}{84}\nu^2 - \frac{504493}{8}\nu + \frac{2812693}{112}}{ (e_r^2 - 1)^{11/2}}
\right){\rm arccos }\left(-\frac{1}{e_r}\right) \nonumber\\
&&  
+ \frac{\frac{312283}{315}\nu^2 - \frac{5265737}{5880}\nu + \frac{20984207}{105840}}{ (e_r^2 - 1)^2} 
+ \frac{\frac{55270210}{3969}\nu^2 - \frac{4854806497}{158760}\nu + \frac{2078715901}{317520}}{ (e_r^2 - 1)^3 }\nonumber\\
&&
+ \frac{\frac{28453511}{756}\nu^2 - \frac{19305805}{216}\nu + \frac{85585279}{3024}}{ (e_r^2 - 1)^4}
+ \frac{\frac{2217065}{84}\nu^2 - \frac{504493}{8}\nu + \frac{2812693}{112}}{ (e_r^2 - 1)^5}
\,.
\eea
\end{widetext}
These functions can be thought as the hyperboliclike counterparts of the EEFF recalled above in the ellipticlike case.
However, they have the meaning of enhancement functions in the limit of large eccentricity only by factoring out the leading order term.
For instance, the Newtonian function $f^{\rm (hyp)}(e_r)$ can be rewritten as
\beq
\label{fhypnewt}
f^{\rm (hyp)}(e_r)=\frac{37\pi}{3(e_r^2 - 1)^{3/2}}\tilde f^{\rm (hyp)}(e_r)
\,,
\eeq
with
\bea
\label{tildefhypnewt}
\tilde f^{\rm (hyp)}(e_r)&=&\left(1+\frac{366e_r^2+59}{37(e_r^2-1)^2}\right)\frac{2}{\pi}{\rm arccos }\left(-\frac{1}{e_r}\right)\nonumber\\
&&
+\frac{2}{\pi}\frac{673e_r^2+602}{111(e_r^2 - 1)^{3/2}}\,,
\eea
so that in the limit $e_r\to\infty$ the latter function approaches unity.
The large-$e_r$ expansion of Eqs. \eqref{fhypnewt} and \eqref{tildefhypnewt} then gives
\beq
f^{\rm (hyp)}(e_r)=\frac{37\pi}{3e_r^3} +\frac{1568}{9e_r^4}+ \frac{281\pi}{2e_r^5}+ \frac{7808}{9e_r^6} 
+O\left(\frac{1}{e_r^7}\right)\,,
\eeq 
and
\beq
\tilde f^{\rm (hyp)}(e_r)=1+\frac{1568}{111\pi e_r}+\frac{366}{37e_r^2}+\frac{5456}{111\pi e_r^3}
+O\left(\frac{1}{e_r^4}\right)\,,
\eeq
respectively.

The remaining multipolar invariants are given by
\begin{widetext}
\bea
I^{\rm (hyp)}_{3[4,4]}&=& 
\frac{\pi\nu^2}{\bar a_r^{9/2}}(1-4\nu)\left[F^{\rm (hyp)}(e_r) +\frac{\eta^2}{\bar a_r}F^{\rm (hyp)}_{\rm 1PN}(e_r;\nu)
+O(\eta^4)\right]
\,,\nonumber\\
J^{\rm (hyp)}_{2[3,3]}&=& 
\frac{\pi\nu^2}{\bar a_r^{9/2}}(1-4\nu)\left[{\mathcal F}^{\rm (hyp)}(e_r) +\frac{\eta^2}{\bar a_r}{\mathcal F}^{\rm (hyp)}_{\rm 1PN}(e_r;\nu)
+O(\eta^4)\right]
\,,\nonumber\\
I^{\rm (hyp)}_{4[5,5]}&=& 
\frac{\pi\nu^2}{\bar a_r^{11/2}}(1-3\nu)^2\left[{\mathfrak F}^{\rm (hyp)}(e_r) 
+O(\eta^2)\right]
\,,\nonumber\\
J^{\rm (hyp)}_{3[4,4]}&=& 
\frac{\pi\nu^2}{\bar a_r^{11/2}}(1-3\nu)^2\left[{\mathbb F}^{\rm (hyp)}(e_r) 
+O(\eta^2)\right]
\,,
\eea
where
\bea
F^{\rm (hyp)}(e_r)&=& 
\left(\frac{165375}{4 (e_r^2 - 1)^{9/2}}+\frac{218715}{4 (e_r^2 - 1)^{7/2}}+\frac{347589}{20 (e_r^2 - 1)^{5/2}}+\frac{15273}{20 (e_r^2 - 1)^{3/2}}\right){\rm arccos }\left(-\frac{1}{e_r}\right)\nonumber\\
&&
+\frac{165375}{4 (e_r^2 - 1)^4}+\frac{81795}{2 (e_r^2 - 1)^3}+\frac{148439}{20 (e_r^2 - 1)^2}
\,,\nonumber\\
{\mathcal F}^{\rm (hyp)}(e_r)&=&
\left(\frac{315}{16 (e_r^2 - 1)^{9/2}}+\frac{455}{16 (e_r^2 - 1)^{7/2}}+\frac{165}{16 (e_r^2 - 1)^{5/2}}+\frac{9}{16 (e_r^2 - 1)^{3/2}}\right){\rm arccos }\left(-\frac{1}{e_r}\right)\nonumber\\
&&
+\frac{315}{16 (e_r^2 - 1)^4}+\frac{175}{8 (e_r^2 - 1)^3}+\frac{229}{48 (e_r^2 - 1)^2}
\,,\nonumber\\
{\mathfrak F}^{\rm (hyp)}(e_r)&=& 
\left(\frac{19022625}{2 (e_r^2 - 1)^{11/2}}+\frac{16795030}{(e_r^2 - 1)^{9/2}}+\frac{62748897}{7 (e_r^2 - 1)^{7/2}}+\frac{50808042}{35 (e_r^2 - 1)^{5/2}}+\frac{2347659}{70 (e_r^2 - 1)^{3/2}}\right){\rm arccos }\left(-\frac{1}{e_r}\right)\nonumber\\
&&
+\frac{19022625}{2 (e_r^2 - 1)^5}+\frac{27249185}{2 (e_r^2 - 1)^4}+\frac{221257987}{42 (e_r^2 - 1)^3}+\frac{4638627}{10 (e_r^2 - 1)^2}
\,,\nonumber\\
{\mathbb F}^{\rm (hyp)}(e_r)&=&
\left(\frac{35189}{8 (e_r^2 - 1)^{11/2}}+\frac{82467}{10 (e_r^2 - 1)^{9/2}}+\frac{57449}{12 (e_r^2 - 1)^{7/2}}+\frac{5273}{6 (e_r^2 - 1)^{5/2}}+\frac{993}{40 (e_r^2 - 1)^{3/2}}\right){\rm arccos }\left(-\frac{1}{e_r}\right)\nonumber\\
&&
+\frac{35189}{8 (e_r^2 - 1)^5}+\frac{813659}{120 (e_r^2 - 1)^4}+\frac{350189}{120 (e_r^2 - 1)^3}+\frac{547187}{1800 (e_r^2 - 1)^2}
\,,
\eea
and PN corrections
\bea
F_{\rm 1PN}^{\rm (hyp)}(e_r;\nu)&=&
\left(\frac{\frac{5590935 \nu }{8}-\frac{16383519}{16}}{(e_r^2 - 1)^{11/2}}+\frac{\frac{9867179 \nu }{8}-\frac{67358543}{40}}{(e_r^2 - 1)^{9/2}}+\frac{\frac{5219637 \nu}{8}-\frac{3088899}{4}}{(e_r^2 - 1)^{7/2}}\right.\nonumber\\
&&\left.
+\frac{\frac{4076289 \nu }{40}-\frac{3204669}{40}}{(e_r^2 - 1)^{5/2}}+\frac{\frac{20259 \nu }{10}+\frac{75807}{80}}{(e_r^2 - 1)^{3/2}}\right){\rm arccos }\left(-\frac{1}{e_r}\right)\nonumber\\
&&
+\frac{\frac{5590935 \nu }{8}-\frac{16383519}{16}}{(e_r^2 - 1)^5}+\frac{\frac{4001767 \nu }{4}-\frac{107411221}{80}}{(e_r^2 - 1)^4}+\frac{\frac{9146293 \nu}{24}-\frac{99767411}{240}}{(e_r^2 - 1)^3}+\frac{\frac{312637 \nu }{10}-\frac{37015357}{2800}}{(e_r^2 - 1)^2}
\,,\nonumber\\
{\mathcal F}_{\rm 1PN}^{\rm (hyp)}(e_r;\nu)&=&
\left(\frac{\frac{12105 \nu }{64}-\frac{125355}{256}}{(e_r^2 - 1)^{11/2}}+\frac{\frac{11249 \nu }{32}-\frac{60333}{64}}{(e_r^2 - 1)^{9/2}}+\frac{\frac{3155 \nu}{16}-\frac{69445}{128}}{(e_r^2 - 1)^{7/2}}+\frac{\frac{7269 \nu }{224}-\frac{39343}{448}}{(e_r^2 - 1)^{5/2}}+\frac{\frac{261 \nu }{448}-\frac{1845}{1792}}{(e_r^2 - 1)^{3/2}}\right){\rm arccos }\left(-\frac{1}{e_r}\right)\nonumber\\
&&
+\frac{\frac{12105 \nu }{64}-\frac{125355}{256}}{(e_r^2 - 1)^5}+\frac{\frac{18463 \nu }{64}-\frac{199547}{256}}{(e_r^2 - 1)^4}+\frac{\frac{22625 \nu}{192}-\frac{83517}{256}}{(e_r^2 - 1)^3}+\frac{\frac{67253 \nu }{6720}-\frac{687077}{26880}}{(e_r^2 - 1)^2}
\,.
\eea

\subsubsection{Angular momentum EEFF}

We list below the multipolar invariants entering the angular momentum flux \eqref{flux2PNhyp}
\bea
I^{*\,\rm (hyp)}_{z\,2[2,3]}&=& 
\frac{\pi\nu^2}{\bar a_r^2}(1-4\nu)\left[f^{*\,\rm (hyp)}(e_r) +\frac{\eta^2}{\bar a_r}f^{*\,\rm (hyp)}_{\rm 1PN}(e_r;\nu)
+\frac{\eta^4}{\bar a_r^2}f^{*\,\rm (hyp)}_{\rm 2PN}(e_r;\nu)\right]
\,,\nonumber\\
I^{*\,\rm (hyp)}_{z\,3[3,4]}&=& 
\frac{\pi\nu^2}{\bar a_r^3}(1-4\nu)\left[F^{*\,\rm (hyp)}(e_r) +\frac{\eta^2}{\bar a_r}F^{*\,\rm (hyp)}_{\rm 1PN}(e_r;\nu)
+O(\eta^4)\right]
\,,\nonumber\\
J^{*\,\rm (hyp)}_{z\,2[2,3]}&=& 
\frac{\pi\nu^2}{\bar a_r^3}(1-4\nu)\left[{\mathcal F}^{*\,\rm (hyp)}(e_r) +\frac{\eta^2}{\bar a_r}{\mathcal F}^{*\,\rm (hyp)}_{\rm 1PN}(e_r;\nu)
+O(\eta^4)\right]
\,,\nonumber\\
I^{*\,\rm (hyp)}_{z\,4[4,5]}&=& 
\frac{\pi\nu^2}{\bar a_r^4}(1-3\nu)^2\left[{\mathfrak F}^{*\,\rm (hyp)}(e_r) 
+O(\eta^2)\right]
\,,\nonumber\\
J^{*\,\rm (hyp)}_{z\,3[3,4]}&=& 
\frac{\pi\nu^2}{\bar a_r^4}(1-3\nu)^2\left[{\mathbb F}^{*\,\rm (hyp)}(e_r) 
+O(\eta^2)\right]
\,,
\eea
where
\bea
f^{*\,\rm (hyp)}(e_r)&=& 
\left(\frac{60}{(e_r^2 - 1)^2}+\frac{28}{e_r^2 - 1}\right){\rm arccos }\left(-\frac{1}{e_r}\right)
+\frac{60}{(e_r^2 - 1)^{3/2}}+\frac{8}{(e_r^2 - 1)^{1/2}}
\,,\nonumber\\
F^{*\,\rm (hyp)}(e_r)&=& 
\left(\frac{9905}{2 (e_r^2 - 1)^3}+\frac{4389}{(e_r^2 - 1)^2}+\frac{5301}{10 e_r^2 - 1}\right){\rm arccos }\left(-\frac{1}{e_r}\right)
+\frac{9905}{2 (e_r^2 - 1)^{5/2}}+\frac{16429}{6 (e_r^2 - 1)^{3/2}}+\frac{288}{5 (e_r^2 - 1)^{1/2}}
\,,\nonumber\\
{\mathcal F}^{*\,\rm (hyp)}(e_r)&=&
\left(\frac{35}{16 (e_r^2 - 1)^3}+\frac{15}{8 (e_r^2 - 1)^2}+\frac{3}{16 e_r^2 - 1}\right){\rm arccos }\left(-\frac{1}{e_r}\right)
+\frac{35}{16 (e_r^2 - 1)^{5/2}}+\frac{55}{48 (e_r^2 - 1)^{3/2}}
\,,\nonumber\\
{\mathfrak F}^{*\,\rm (hyp)}(e_r)&=& 
\left(\frac{787775}{(e_r^2 - 1)^4}+\frac{1033545}{(e_r^2 - 1)^3}+\frac{2287227}{7 (e_r^2 - 1)^2}+\frac{107829}{7 e_r^2 - 1}\right){\rm arccos }\left(-\frac{1}{e_r}\right)\nonumber\\
&&
+\frac{787775}{(e_r^2 - 1)^{7/2}}+\frac{2312860}{3 (e_r^2 - 1)^{5/2}}+\frac{978507}{7 (e_r^2 - 1)^{3/2}}+\frac{4608}{7 (e_r^2 - 1)^{1/2}}
\,,\nonumber\\
{\mathbb F}^{*\,\rm (hyp)}(e_r)&=&
\left(\frac{385}{(e_r^2 - 1)^4}+\frac{4655}{9 (e_r^2 - 1)^3}+\frac{505}{3 (e_r^2 - 1)^2}+\frac{23}{3 e_r^2 - 1}\right){\rm arccos }\left(-\frac{1}{e_r}\right)\nonumber\\
&&
+\frac{385}{(e_r^2 - 1)^{7/2}}+\frac{3500}{9 (e_r^2 - 1)^{5/2}}+\frac{1969}{27 (e_r^2 - 1)^{3/2}}
\,,
\eea
and PN corrections
\bea
f_{\rm 1PN}^{*\,\rm (hyp)}(e_r;\nu)&=&
\left(\frac{530 \nu -\frac{1415}{3}}{(e_r^2 - 1)^3}+\frac{484 \nu -318}{(e_r^2 - 1)^2}+\frac{\frac{414 \nu }{7}+\frac{475}{21}}{e_r^2 - 1}\right){\rm arccos }\left(-\frac{1}{e_r}\right)\nonumber\\
&&
+\frac{530 \nu -\frac{1415}{3}}{(e_r^2 - 1)^{5/2}}+\frac{\frac{922 \nu }{3}-\frac{1447}{9}}{(e_r^2 - 1)^{3/2}}+\frac{\frac{36\nu }{7}+\frac{184}{7}}{(e_r^2 - 1)^{1/2}}
\,,\nonumber\\
f_{\rm 2PN}^{*\,\rm (hyp)}(e_r;\nu)&=&
\left(\frac{\frac{553145 \nu ^2}{189}-\frac{2574049 \nu}{378}+\frac{2001929}{1512}}{(e_r^2 - 1)^4}+\frac{\frac{231445 \nu ^2}{63}-\frac{1052159 \nu}{126}+\frac{133115}{168}}{(e_r^2 - 1)^3}+\frac{\frac{444475 \nu ^2}{441}-\frac{1783919 \nu}{882}-\frac{509945}{3528}}{(e_r^2 - 1)^2}\right.\nonumber\\
&&\left.
+\frac{\frac{869 \nu ^2}{49}+\frac{29263 \nu}{294}+\frac{12317}{392}}{e_r^2 - 1}\right){\rm arccos }\left(-\frac{1}{e_r}\right)\nonumber\\
&&
+\frac{\frac{553145 \nu ^2}{189}-\frac{2574049 \nu}{378}+\frac{2001929}{1512}}{(e_r^2 - 1)^{7/2}}+\frac{\frac{1529860 \nu ^2}{567}-\frac{3447691
\nu }{567}+\frac{199022}{567}}{(e_r^2 - 1)^{5/2}}+\frac{\frac{162718 \nu ^2}{441}-\frac{278169 \nu }{490}-\frac{8106667}{52920}}{(e_r^2 - 1)^{3/2}}\nonumber\\
&&
+\frac{-\frac{295 \nu ^2}{147}+\frac{3727 \nu }{147}+\frac{9712}{147}}{(e_r^2 - 1)^{1/2}}
\,,\nonumber\\
F_{\rm 1PN}^{*\,\rm (hyp)}(e_r;\nu)&=&
\left(\frac{\frac{392665 \nu }{6}-\frac{277564}{3}}{(e_r^2 - 1)^4}+\frac{84350 \nu-\frac{217217}{2}}{(e_r^2 - 1)^3}+\frac{\frac{252979 \nu}{10}-\frac{120613}{5}}{(e_r^2 - 1)^2}+\frac{\frac{4623 \nu }{5}+\frac{7131}{10}}{e_r^2 - 1}\right){\rm arccos }\left(-\frac{1}{e_r}\right)\nonumber\\
&&
+\frac{\frac{392665 \nu }{6}-\frac{277564}{3}}{(e_r^2 - 1)^{7/2}}+\frac{\frac{1125635 \nu}{18}-\frac{1399825}{18}}{(e_r^2 - 1)^{5/2}}+\frac{\frac{154051 \nu
}{15}-\frac{192721}{30}}{(e_r^2 - 1)^{3/2}}+\frac{\frac{48 \nu }{5}+\frac{1344}{5}}{(e_r^2 - 1)^{1/2}}
\,,\nonumber\\
{\mathcal F}_{\rm 1PN}^{*\,\rm (hyp)}(e_r;\nu)&=&
\left(\frac{15 \nu -\frac{1971}{64}}{(e_r^2 - 1)^4}+\frac{\frac{305 \nu}{16}-\frac{2875}{64}}{(e_r^2 - 1)^3}+\frac{\frac{303 \nu}{56}-\frac{6819}{448}}{(e_r^2 - 1)^2}+\frac{\frac{15 \nu }{112}-\frac{219}{448}}{e_r^2 - 1}\right){\rm arccos }\left(-\frac{1}{e_r}\right)\nonumber\\
&&
+\frac{15 \nu -\frac{1971}{64}}{(e_r^2 - 1)^{7/2}}+\frac{\frac{225 \nu}{16}-\frac{1109}{32}}{(e_r^2 - 1)^{5/2}}+\frac{\frac{691 \nu }{336}-\frac{43051}{6720}}{(e_r^2 - 1)^{3/2}}
\,.
\eea

\end{widetext}

\subsection{Mapping observables from unbound to bound orbits}

The possibility to pass from elliptic to hyperbolic expressions (for local quantities expressed in PN sense) was first noticed by K\"alin and Porto \cite{Kalin:2019rwq,Kalin:2019inp} (see also Ref. \cite{Cho:2021arx}), who established a relation between the scattering angle for unbound orbits and the periastron advance for bound orbits by analytic continuation.
Their argument was further investigated in Ref. \cite{Saketh:2021sri}, where the mapping was extended to general observables according to their parity behavior under the exchange $j\to-j$. 

Consider, for instance, the quantities $I_{2[3,3]}^{\rm (ell)}(\bar E, j)$ and $I_{2[3,3]}^{\rm (hyp)}(\bar E, j)$, whose expressions are listed in Table \ref{tab:tableI2_33}.
It is easy to see that they satisfy the relation
\beq
\label{enmap}
I_{2[3,3]}^{\rm (ell)}(\bar E, j)=I_{2[3,3]}^{\rm (hyp)}(\bar E, j)-I_{2[3,3]}^{\rm (hyp)}(\bar E, -j)\,,
\eeq
by using the following odd-type extension of the ${\rm arccos}={\rm arccos}\left(-\frac1{\sqrt{1+2\bar Ej^2}}\right)$ term (see Eq. (4.20) in Ref. \cite{Saketh:2021sri}) to
\beq
\overline{\rm arccos}={\rm arccos}\left(-\frac1{j\sqrt{1/j^2+2\bar E}}\right)\,,
\eeq
taking also into account that ${\rm arccos}(-x)=\pi-{\rm arccos}(x)$, and that the coefficients $A$, $B$, $C$ of the $\overline{\rm arccos}$ terms are odd functions of $j$, whereas the coefficients $A_2$ and $B_2$ are even functions of $j$.
This is equivalent to replace ${\rm arccos}\left(-\frac1{\sqrt{1+2\bar Ej^2}}\right)\to \pi$, then adopting the K\"alin-Porto prescription of killing all even in $j$ functions $A_2=0=B_2$ (the functions $A\,\overline{\rm arccos}$, $B\,\overline{\rm arccos}$ and $C\,\overline{\rm arccos}$ being all even too).  
 
The multipolar invariants entering the angular momentum loss have different parity with respect to the corresponding energy-related quantities.
Consider, for instance, $I_{z\,2[2,3]}^{*\,\rm (ell)}(\bar E, j)$ and $I_{z\,2[2,3]}^{*\,\rm (hyp)}(\bar E, j)$, whose expressions are listed in Table \ref{tab:tableIstar2_23}.
It is easy to see that in this case
\beq
\label{angmommap}
I_{z\,2[2,3]}^{*\,\rm (ell)}(\bar E, j)=I_{z\,2[2,3]}^{*\,\rm (hyp)}(\bar E, j)+I_{z\,2[2,3]}^{*\,\rm (hyp)}(\bar E, -j)\,.
\eeq

The properties \eqref{enmap} and \eqref{angmommap} are also valid for all invariants computed here.


\begin{table*}  
\caption{\label{tab:tableI2_33}  $I_{2[3,3]}$: ellipticlike vs hyperboliclike contributions. }
\begin{ruledtabular}
\begin{tabular}{ll}
$I_{2[3,3]}^{\rm (ell)}(\bar E, j)$  & $\pi\nu^2\left[A+B\eta^2+C\eta^4\right] $\\
$A$ &$\frac{74}{ 3 j^3} (-2\bar E)^2-\frac{244}{j^5}(-2\bar E)+\frac{850}{ 3 j^7}$\\
$B$ &$\frac{(-\frac{1004}{21}\nu-\frac{51}{7})}{ j^3}(-2\bar E)^3+\frac{(\frac{7283}{7}\nu+\frac{538}{7})}{ j^5} (-2\bar E)^2+\frac{(-\frac{10015}{3}\nu-\frac{3980}{3})}{j^7} (-2\bar E)+\frac{(\frac{7910}{3}\nu+2233)}{j^9}$\\
$C$ &$\frac{(\frac{35981}{784}+\frac{16229}{1176}\nu^2-\frac{20555}{392}\nu)}{j^3} (-2\bar E)^4
+\frac{(-\frac{1118539}{1764}-\frac{291511}{294}\nu^2+\frac{1007617}{294}\nu)}{j^5} (-2\bar E)^3
+\frac{(\frac{7070135}{882}\nu^2-\frac{32607265}{1764}\nu+\frac{8304755}{3528})}{j^7} (-2\bar E)^2$\\
&$
+\frac{(\frac{604436}{27}\nu-\frac{1438393}{108}-\frac{495377}{27}\nu^2)}{j^9} (-2\bar E)
+\frac{(-\frac{69883}{24}\nu+\frac{7037855}{336}+\frac{258245}{21}\nu^2)}{j^{11}}$\\
\hline
$I_{2[3,3]}^{\rm (hyp)}(\bar E, j)$  &$\nu^2\left\{A{\rm arccos}\left(-\frac1{\sqrt{1+2\bar Ej^2}}\right)+A_2\sqrt{2\bar E}+\left[B{\rm arccos}\left(-\frac1{\sqrt{1+2\bar Ej^2}}\right)+B_2\frac{\sqrt{2\bar E}}{1+2\bar Ej^2}\right]\eta^2\right.$\\
 & $\left.+\left[C{\rm arccos}\left(-\frac1{\sqrt{1+2\bar Ej^2}}\right)+C_2\frac{\sqrt{2\bar E}}{(1+2\bar Ej^2)^2}\right]\eta^4\right\} $\\
$A_2$ & $\frac{1346}{ 9 j^4} (2\bar E)+\frac{850}{ 3 j^6}$\\
$B_2$ &$\frac{(\frac{28160}{63}\nu+\frac{2909}{315})}{j^2} (2\bar E)^3
+\frac{(\frac{20401}{7}\nu+\frac{209024}{315})}{j^4} (2\bar E)^2
+\frac{(\frac{8446}{3}+\frac{45865}{9}\nu)}{j^6} (2\bar E)+\frac{(\frac{7910}{3}\nu+2233)}{j^8}$\\ 
$C_2$ &$ (\frac{35124473}{105840}-\frac{18370789}{17640}\nu+\frac{667909}{2520}\nu^2) (2\bar E)^5
+\frac{(-\frac{317325973}{22680}\nu+\frac{770937919}{158760}\nu^2+\frac{595470733}{317520})}{j^2} (2\bar E)^4$\\
&$
+\frac{(\frac{1597862467}{158760}-\frac{3647746597}{79380}\nu+\frac{460401199}{19845}\nu^2)}{j^4} (2\bar E)^3
+\frac{(\frac{1134263257}{31752}+\frac{358428137}{7938}\nu^2-\frac{4552336937}{79380}\nu)}{j^6} (2\bar E)^2$\\
&$
+\frac{(-\frac{5883733}{216}\nu+\frac{7341314}{189}\nu^2+\frac{145842829}{3024})}{j^8} (2\bar E)
+\frac{(-\frac{69883}{24}\nu+\frac{7037855}{336}+\frac{258245}{21}\nu^2)}{j^{10}}$\\
\end{tabular}
\end{ruledtabular}
\end{table*}


\begin{table*}  
\caption{\label{tab:tableIstar2_23}  $I^*_{z\,2[2,3]}$: ellipticlike vs hyperboliclike contributions. }
\begin{ruledtabular}
\begin{tabular}{ll}
$I_{z\,2[2,3]}^{*\,\rm (ell)}(\bar E, j)$  & $\pi\nu^2\left[A+B\eta^2+C\eta^4\right] $\\
$A$ &$\frac{28}{j^2}(2\bar E)+\frac{60}{j^4}$\\
$B$ &$ \frac{(\frac{622}{21}+\frac{267}{7}\nu)}{j^2}(2\bar E)^2+\frac{(90+336\nu)}{j^4} (2\bar E)+\frac{(410\nu+\frac{745}{3})}{j^6}$\\
$C$ &$\frac{(\frac{106}{49}\nu^2-\frac{983}{392}-\frac{391}{98}\nu)}{j^2} (2\bar E)^3
+\frac{(-\frac{657365}{3528}-\frac{458131}{441}\nu+\frac{132625}{441}\nu^2)}{j^4} (2\bar E)^2
+\frac{(-\frac{33205}{168}+\frac{94735}{63}\nu^2-\frac{388265}{126}\nu)}{j^6} (2\bar E)$\\
&$
+\frac{(-\frac{519619}{378}\nu+\frac{1865849}{1512}+\frac{286655}{189}\nu^2)}{j^8} $\\
\hline
$I_{z\,2[2,3]}^{*\,\rm (hyp)}(\bar E, j)$  &$\nu^2\left\{A{\rm arccos}\left(-\frac1{\sqrt{1+2\bar Ej^2}}\right)+A_2\sqrt{2\bar E}+\left[B{\rm arccos}\left(-\frac1{\sqrt{1+2\bar Ej^2}}\right)+B_2\frac{\sqrt{2\bar E}}{1+2\bar Ej^2}\right]\eta^2\right.$\\
 & $\left.+\left[C{\rm arccos}\left(-\frac1{\sqrt{1+2\bar Ej^2}}\right)+C_2\frac{\sqrt{2\bar E}}{(1+2\bar Ej^2)^2}\right]\eta^4\right\} $\\
$A_2$&$ \frac{8}{j} (2\bar E)+\frac{60}{j^3} $\\
$B_2$ &$(\frac{29}{7}\nu+\frac{93}{7}) j (2\bar E)^3+\frac{(\frac{4357}{21}\nu+\frac{3560}{63})}{j} (2\bar E)^2+\frac{(\frac{2300}{9}+\frac{1828}{3}\nu)}{j^3} (2\bar E)+\frac{(410\nu+\frac{745}{3})}{j^5} $\\
$C_2$&$ (-\frac{1591}{2352}\nu^2-\frac{1079}{2352}+\frac{11357}{1176}\nu) j^3 (2\bar E)^5
+(\frac{618451}{7056}\nu^2-\frac{7495169}{105840}-\frac{2137883}{5880}\nu) j (2\bar E)^4$\\
&$+\frac{(-\frac{40681301}{79380}-\frac{129949573}{39690}\nu+\frac{4722409}{3969}\nu^2)}{j} (2\bar E)^3+\frac{(\frac{14351336}{3969}\nu^2-\frac{274169579}{39690}\nu+\frac{5652061}{39690})}{j^3} (2\bar E)^2
$\\
&$
+\frac{(\frac{4216355}{2268}+\frac{2285890}{567}\nu^2-\frac{3046240}{567}\nu)}{j^5} (2\bar E)
+\frac{(-\frac{519619}{378}\nu+\frac{1865849}{1512}+\frac{286655}{189}\nu^2)}{j^7} $
\end{tabular}
\end{ruledtabular}
\end{table*}

\subsection{Tail part}

The energy and angular momentum tails \eqref{F_G_tail} at the 1PN fractional accuracy read
\bea
(\Delta E)_{\rm tail}^{\rm (hyp)} &=&  
\frac{G^2{\mathcal M}}{c^8}  \int_0^\infty  d\omega\, \left[\frac{2}{5}\omega^7 \kappa^{I_2}(\omega)\right.\nonumber\\
&&\left.
+\eta^2\left(\frac{2}{189}\omega^9 \kappa^{I_3}(\omega)+\frac{32}{45}\omega^7 \kappa^{J_2}(\omega)\right)\right]
\,,\nonumber\\
(\Delta J_i)_{\rm tail}^{\rm (hyp)}&=&
\frac{G^2{\mathcal M}}{c^8} \int_0^\infty d\omega\, \left[\frac{2}{5}\omega^6 \kappa^{I_2}_i(\omega)\right.\nonumber\\
&&\left.
+\eta^2\left(\frac{1}{63}\omega^8 \kappa^{I_3}_i(\omega)+\frac{32}{45}\omega^6 \kappa^{J_2}_i(\omega)\right)\right]
\,,\nonumber\\
\eea
where we have introduced the notation \cite{Bini:2021qvf}
\bea
\label{kappatensdef}
\kappa^{X_l}_{ab}(\omega)&=&\hat X_{aL-1}(\omega) \hat X_{bL-1} (-\omega)=\kappa^{X_l}_{ba}(-\omega)
\,,\nonumber\\ 
\kappa^{X_l}(\omega)&=&{\rm Tr}[\kappa^{X_l}_{ab}(\omega)]
\,,\nonumber\\
\kappa^{X_l}_i(\omega)&=&2i \epsilon_{iab} \kappa^{X_l}_{ab}(\omega)
\,.
\eea

The Newtonian terms have been computed in Ref. \cite{Bini:2021qvf} in a large-eccentricity expansion.
We push the calculation to the 1PN fractional accuracy by using the tools already developed in previous works (see, e.g., Ref. \cite{Bini:2021jmj}). 
The main difficulty going beyond the Newtonian level is that in some cases the Fourier transform of multipole moments cannot be directly evaluated. 
Nevertheless, one can always obtain analytical results by integrating over the frequencies first.

We find
\begin{widetext}
\bea
\label{1PNtails}
(\Delta E)_{\rm tail}^{\rm (hyp)}&=& \nu^2 \left\{
\frac{\frac{3136}{45} p_\infty^6
+\left(\frac{1216}{105}-\frac{2848 \nu}{15}\right) p_\infty^8}{j^4}\right.\nonumber\\
&&
+\pi\frac{\frac{297}{20} \pi^2 p_\infty^5
+\left(\frac{9216}{35}-\frac{24993 \pi^2}{1120}-\frac{15291 \pi^2 \nu}{280}\right) p_\infty^7}{j^5}\nonumber\\
&&
+\frac{\left(\frac{9344}{45}+\frac{88576 \pi ^2}{675}\right) p_\infty^4
+\left[\left(-\frac{52928}{45}-\frac{3014912 \pi^2}{4725}\right) \nu+ \frac{2898 \zeta(3)}{5}+\frac{56708}{105}+\frac{1024 \pi^2}{135}\right]p_\infty^6 }{j^6}\nonumber\\
&&\left.
+\pi\frac{\left(\frac{1579 \pi ^2}{3}-\frac{2755 \pi ^4}{64}\right) p_\infty^3
+\left[\left(\frac{30285 \pi ^4}{112}-\frac{23514\pi ^2}{7}\right) \nu +\frac{210176 }{225}-\frac{13138915 \pi ^2}{7392}+\frac{689985 \pi^4}{3584}\right] p_\infty^5}{j^7}+O\left(\frac1{j^8}
\right)\right\}
\,,\nonumber\\
(\Delta J_z)_{\rm tail}^{\rm (hyp)}&=& \nu^2 \left\{ 
\frac{\frac{448}{5} p_\infty^4
+\left(-\frac{17600}{63}\nu  + \frac{1184}{21}\right) p_\infty^6}{j^3} \right.\nonumber\\
&&
+\pi \frac{\frac{69}{5}\pi^2p_\infty^3  
+\left(- \frac{2232}{35}\nu \pi^2   -\frac{1305}{112}\pi^2 + \frac{7488}{25}\right)p_\infty^5}{j^4}\nonumber\\ 
&&
+\frac{\left(\frac{128}{15} +  \frac{4352}{45}\pi^2 \right) p_\infty^2
+\left[\left(-\frac{30592}{63} -  \frac{156416}{6615}\pi^2 \right)\nu  +  \frac{4116}{5}\zeta(3) -  \frac{130688}{6615}\pi^2  + \frac{147064}{315} \right]p_\infty^4}{j^5} \nonumber\\
&&\left. 
+ \pi \frac{\left(-\frac{423}{16}\pi^4 + 303\pi^2\right) p_\infty
+\left[\left(\frac{102619}{448}\pi^4 - \frac{57037}{21}\pi^2\right)\nu  +  \frac{163083}{1792}\pi^4 - \frac{18227}{28}\pi^2 + \frac{32}{15}  \right]p_\infty^3}{j^6}
+O\left(\frac1{j^7}\right)
\right\}
\,.
\eea
\end{widetext}
The fractional 1PN energy tail agrees with the recent result by Cho \cite{Cho:2022pqy} through a combination of analytic and numerical techniques, here obtained fully analytically.
The 1PN angular momentum tail is instead a new result.

\section{Evolution of orbital elements}

The knowledge of the gravitational wave energy and angular momentum fluxes allows for studying how the orbital elements evolve under the gravitational radiation reaction.
This is done simply by differentiating the orbital elements with respect to time, using then the balance equations to equate the decreases of energy and angular momentum to the corresponding averaged fluxes.
In the case of ellipticlike orbits the slow secular evolution under gravitational radiation reaction at 3PN order has been computed in Ref. \cite{Arun:2009mc}. 

We discuss below the averaged adiabatic evolution of the orbital elements $\bar n$, $k$, $\bar a_r$ and $e_r$ in the case of hyperboliclike motion at the 2.5PN order, by separating out the contributions due to the instantaneous and hereditary terms in the fluxes.
Concerning the instantaneous part we find
\begin{widetext}
\bea
\left\langle\frac{d\bar n}{dt}\right\rangle^{\rm (hyp)}_{\rm inst}&=&\frac{\nu}{\bar a_r^4}\left\{
\left(-\frac{170}{(e_r^2-1)^{7/2}}-\frac{732}{5 (e_r^2-1)^{5/2}}-\frac{74}{5(e_r^2-1)^{3/2}}\right){\rm arccos }\left(-\frac{1}{e_r}\right)
-\frac{170}{(e_r^2-1)^3}-\frac{1346}{15 (e_r^2-1)^2}\right.\nonumber\\
&&
+\frac{\eta^2}{\bar a_r}\left[
\left(\frac{532 \nu +\frac{31059}{20}}{(e_r^2-1)^{9/2}}+\frac{\frac{2057 \nu }{3}+\frac{4047}{4}}{(e_r^2-1)^{7/2}}+\frac{\frac{1102 \nu}{5}-\frac{43341}{140}}{(e_r^2-1)^{5/2}}+\frac{\frac{37 \nu }{3}-\frac{11717}{140}}{(e_r^2-1)^{3/2}}\right){\rm arccos }\left(-\frac{1}{e_r}\right)\right.\nonumber\\
&&\left.
+\frac{532 \nu +\frac{31059}{20}}{(e_r^2-1)^4}+\frac{\frac{1525 \nu}{3}+\frac{4941}{10}}{(e_r^2-1)^3}+\frac{\frac{4421 \nu }{45}-\frac{235367}{700}}{(e_r^2-1)^2}
\right]\nonumber\\
&&
+\frac{\eta^4}{\bar a_r^2}\left[
\left(\frac{-378 \nu ^2-\frac{207469 \nu }{10}-\frac{2678867}{1680}}{(e_r^2-1)^{11/2}}+\frac{-\frac{2590 \nu ^2}{3}-\frac{3282923 \nu}{120}+\frac{6214991}{1080}}{(e_r^2-1)^{9/2}}+\frac{-\frac{8305 \nu ^2}{12}-\frac{6316991 \nu }{840}+\frac{414997}{90}}{(e_r^2-1)^{7/2}}\right.\right.\nonumber\\
&&\left.
+\frac{-\frac{2101 \nu^2}{10}+\frac{131267 \nu }{280}-\frac{1299329}{840}}{(e_r^2-1)^{5/2}}+\frac{-\frac{259 \nu ^2}{20}+\frac{4769 \nu }{56}-\frac{763541}{1680}}{(e_r^2-1)^{3/2}}\right){\rm arccos }\left(-\frac{1}{e_r}\right)\nonumber\\
&&
+\frac{-378 \nu ^2-\frac{207469 \nu }{10}-\frac{2678867}{1680}}{(e_r^2-1)^5}+\frac{-\frac{2212 \nu ^2}{3}-\frac{2453047 \nu}{120}+\frac{19009295}{3024}}{(e_r^2-1)^4}+\frac{-\frac{86383 \nu ^2}{180}-\frac{8033687 \nu }{3150}+\frac{538413661}{226800}}{(e_r^2-1)^3}\nonumber\\
&&\left.\left.
+\frac{-\frac{17653\nu ^2}{180}+\frac{5895007 \nu }{12600}-\frac{33420491}{19600}}{(e_r^2-1)^2}
\right]
\right\}
\,,\nonumber\\
\left\langle\frac{dk}{dt}\right\rangle^{\rm (hyp)}_{\rm inst}&=&\frac{\nu}{\bar a_r^{7/2}}\eta^2\left\{
\left(\frac{144}{(e_r^2-1)^{7/2}}+\frac{336}{5 (e_r^2-1)^{5/2}}\right){\rm arccos }\left(-\frac{1}{e_r}\right)
+\frac{144}{(e_r^2-1)^3}+\frac{96}{5 (e_r^2-1)^2}\right.\nonumber\\
&&
+\frac{\eta^2}{\bar a_r}\left[
\left(\frac{148-986 \nu }{(e_r^2-1)^{9/2}}+\frac{-\frac{2436 \nu }{5}-384}{(e_r^2-1)^{7/2}}+\frac{\frac{94 \nu}{5}-\frac{1716}{35}}{(e_r^2-1)^{5/2}}\right){\rm arccos }\left(-\frac{1}{e_r}\right)\right.\nonumber\\
&&\left.\left.
+\frac{148-986 \nu }{(e_r^2-1)^4}+\frac{-\frac{2378 \nu }{15}-\frac{1300}{3}}{(e_r^2-1)^3}+\frac{\frac{48 \nu}{5}+\frac{1224}{35}}{(e_r^2-1)^2}
\right]
\right\}
\,,\nonumber\\
\left\langle\frac{d\bar a_r}{dt}\right\rangle^{\rm (hyp)}_{\rm inst}&=&\frac{\nu}{\bar a_r^{3/2}}\left\{
\left(\frac{340}{3 (e_r^2-1)^{7/2}}+\frac{488}{5 (e_r^2-1)^{5/2}}+\frac{148}{15 (e_r^2-1)^{3/2}}\right){\rm arccos }\left(-\frac{1}{e_r}\right)
+\frac{340}{3 (e_r^2-1)^3}+\frac{2692}{45 (e_r^2-1)^2}\right.\nonumber\\
&&
+\frac{\eta^2}{\bar a_r}\left[
\left(\frac{-\frac{1064 \nu }{3}-\frac{10353}{10}}{(e_r^2-1)^{9/2}}+\frac{-\frac{1088 \nu }{3}-\frac{3049}{2}}{(e_r^2-1)^{7/2}}+\frac{-\frac{328 \nu}{5}-\frac{36793}{70}}{(e_r^2-1)^{5/2}}-\frac{3823}{210 (e_r^2-1)^{3/2}}\right){\rm arccos }\left(-\frac{1}{e_r}\right)\right.\nonumber\\
&&\left.
++\frac{-\frac{1064 \nu }{3}-\frac{10353}{10}}{(e_r^2-1)^4}+\frac{-\frac{2200 \nu}{9}-\frac{5897}{5}}{(e_r^2-1)^3}+\frac{-\frac{704 \nu }{45}-\frac{235733}{1050}}{(e_r^2-1)^2}
\right]\nonumber\\
&&
+\frac{\eta^4}{\bar a_r^2}\left[
\left(\frac{252 \nu ^2+\frac{207469 \nu }{15}+\frac{2678867}{2520}}{(e_r^2-1)^{11/2}}+\frac{280 \nu ^2+\frac{990857 \nu}{45}+\frac{1131976}{405}}{(e_r^2-1)^{9/2}}+\frac{60 \nu ^2+\frac{303151 \nu }{35}+\frac{1466927}{540}}{(e_r^2-1)^{7/2}}\right.\right.\nonumber\\
&&\left.
+\frac{\frac{49243 \nu}{105}+\frac{245591}{315}}{(e_r^2-1)^{5/2}}+\frac{41567}{2520 (e_r^2-1)^{3/2}}\right){\rm arccos }\left(-\frac{1}{e_r}\right)\nonumber\\
&&
+\frac{252 \nu ^2+\frac{207469 \nu }{15}+\frac{2678867}{2520}}{(e_r^2-1)^5}+\frac{196 \nu ^2+\frac{783388 \nu}{45}+\frac{11070811}{4536}}{(e_r^2-1)^4}+\frac{\frac{256 \nu ^2}{15}+\frac{19315937 \nu }{4725}+\frac{679540139}{340200}}{(e_r^2-1)^3}
\nonumber\\
&&\left.\left.
+\frac{\frac{2192 \nu}{225}+\frac{24808963}{88200}}{(e_r^2-1)^2}
\right]
\right\}
\,,\nonumber\\
\eea
\bea
\left\langle\frac{de_r}{dt}\right\rangle^{\rm (hyp)}_{\rm inst}&=&\frac{\nu}{e_r\bar a_r^{5/2}}\left\{
\left(-\frac{170}{3 (e_r^2-1)^{5/2}}-\frac{364}{5 (e_r^2-1)^{3/2}}-\frac{242}{15 (e_r^2-1)^{1/2}}\right){\rm arccos }\left(-\frac{1}{e_r}\right)
-\frac{170}{3 (e_r^2-1)^2}-\frac{2426}{45 (e_r^2-1)}-\frac{16}{5}\right.\nonumber\\
&&
+\frac{\eta^2}{\bar a_r}\left[
\left(\frac{\frac{532 \nu }{3}+\frac{10353}{20}}{(e_r^2-1)^{7/2}}+\frac{\frac{736 \nu }{3}+\frac{10945}{12}}{(e_r^2-1)^{5/2}}+\frac{68 \nu
   +\frac{56981}{140}}{(e_r^2-1)^{3/2}}+\frac{5281}{420 (e_r^2-1)^{1/2}}\right){\rm arccos }\left(-\frac{1}{e_r}\right)\right.\nonumber\\
&&\left.
+\frac{\frac{532 \nu }{3}+\frac{10353}{20}}{(e_r^2-1)^3}+\frac{\frac{1676 \nu
   }{9}+\frac{11093}{15}}{(e_r^2-1)^2}+\frac{\frac{976 \nu }{45}+\frac{1301009}{6300}}{e_r^2-1}-\frac{288}{35}
\right]\nonumber\\
&&
+\frac{\eta^4}{\bar a_r^2}\left[
\left(\frac{-126 \nu ^2-\frac{207469 \nu }{30}-\frac{2678867}{5040}}{(e_r^2-1)^{9/2}}+\frac{-180 \nu ^2-\frac{390389 \nu}{30}-\frac{730177}{540}}{(e_r^2-1)^{7/2}}+\frac{-54 \nu ^2-\frac{466139 \nu }{70}-\frac{442379}{360}}{(e_r^2-1)^{5/2}}\right.\right.\nonumber\\
&&\left.
+\frac{-\frac{117977 \nu}{210}-\frac{1596299}{3780}}{(e_r^2-1)^{3/2}}-\frac{3391}{240 (e_r^2-1)^{1/2}}\right){\rm arccos }\left(-\frac{1}{e_r}\right)\nonumber\\
&&
+\frac{-126 \nu ^2-\frac{207469 \nu }{30}-\frac{2678867}{5040}}{(e_r^2-1)^4}+\frac{-138 \nu ^2-\frac{481849 \nu}{45}-\frac{17766089}{15120}}{(e_r^2-1)^3}+\frac{-\frac{96 \nu ^2}{5}-\frac{11669489 \nu }{3150}-\frac{200583793}{226800}}{(e_r^2-1)^2}\nonumber\\
&&\left.\left.
+\frac{-\frac{8644 \nu}{525}-\frac{10176053}{58800}}{e_r^2-1}-\frac{808}{45}
\right]
\right\}
\,.
\eea

The corresponding hereditary contributions at the fractional 1PN accuracy are given by
\bea
\left\langle\frac{d\bar n}{dt}\right\rangle^{\rm (hyp)}_{\rm hered}&=&\frac{\nu}{\bar a_r^{11/2}}\left\{
-\frac{3136}{15 e_r^4}
-\frac{891 \pi ^3}{20 e_r^5}
+\frac{-\frac{15616}{15}-\frac{88576 \pi ^2}{225}}{e_r^6}
+\frac{\frac{8265 \pi ^5}{64}-\frac{13523 \pi ^3}{8}}{e_r^7}
\right.\nonumber\\
&&
+\frac{\eta^2}{\bar a_r}\left[
\frac{\frac{9952 \nu}{45}-\frac{9024}{7}}{e_r^4}
+\frac{\frac{9423 \pi ^3 \nu }{140}-\frac{24921 \pi ^3}{224}-\frac{27648 \pi }{35}}{e_r^5}
+\frac{\left(\frac{17024}{9}+\frac{816896 \pi ^2}{945}\right) \nu -\frac{8694 \zeta (3)}{5}-\frac{182272 \pi^2}{225}-\frac{308236}{105}}{e_r^6}\right.\nonumber\\
&&\left.\left.
+\frac{\left(\frac{431269 \pi ^3}{84}-\frac{360425 \pi ^5}{896}\right) \nu -\frac{2069955 \pi ^5}{3584}+\frac{28200311 \pi ^3}{4928}-\frac{2508032\pi }{525}}{e_r^7}
\right]
+O\left(\frac1{e_r^8}\right)
\right\}
\,,\nonumber\\
\left\langle\frac{dk}{dt}\right\rangle^{\rm (hyp)}_{\rm hered}&=&\frac{\nu}{\bar a_r^5}\eta^2\left[
\frac{2688}{5 e_r^6}
+\frac{414 \pi ^3}{5 e_r^7}
+\frac{1664+\frac{8704 \pi ^2}{15}}{e_r^8}
+\frac{\frac{10539 \pi ^3}{5}-\frac{1269 \pi ^5}{8}}{e_r^9}
+O\left(\frac1{e_r^{10}}\right)
\right]
\,,\nonumber\\
\left\langle\frac{d\bar a_r}{dt}\right\rangle^{\rm (hyp)}_{\rm hered}&=&\frac{\nu}{\bar a_r^{11/2}}\left\{
\frac{6272}{45 e_r^4}
+\frac{297 \pi ^3}{10 e_r^5}
+\frac{\frac{31232}{45}+\frac{177152 \pi ^2}{675}}{e_r^6}
+\frac{\frac{13523 \pi ^3}{12}-\frac{2755 \pi ^5}{32}}{e_r^7}
\right.\nonumber\\
&&
+\frac{\eta^2}{\bar a_r}\left[
\frac{-\frac{1408\nu }{45}-\frac{3904}{21}}{e_r^4}
+\frac{-\frac{2817 \pi ^3 \nu }{140}-\frac{16641 \pi ^3}{112}+\frac{18432 \pi }{35}}{e_r^5}
+\frac{\left(-\frac{2048}{3}-\frac{22528 \pi ^2}{63}\right) \nu +\frac{5796 \zeta (3)}{5}-\frac{964096 \pi^2}{675}-\frac{1023208}{315}}{e_r^6}
\right.\nonumber\\
&&\left.\left.
+\frac{\left(\frac{1375 \pi ^5}{7}-\frac{417257 \pi ^3}{168}\right) \nu +\frac{1847085 \pi ^5}{1792}-\frac{90676571 \pi ^3}{7392}+\frac{5016064 \pi}{1575}}{e_r^7}
\right]
+O\left(\frac1{e_r^8}\right)
\right\}
\,,\nonumber\\
\left\langle\frac{de_r}{dt}\right\rangle^{\rm (hyp)}_{\rm hered}&=&\frac{\nu}{\bar a_r^{11/2}}\left\{
-\frac{7168}{45 e_r^3}
-\frac{573 \pi ^3}{20 e_r^4}
+\frac{-\frac{5632}{15}-\frac{153856 \pi ^2}{675}}{e_r^5}
+\frac{\frac{4447 \pi ^5}{64}-\frac{104677 \pi ^3}{120}}{e_r^6}
\right.\nonumber\\
&&
+\frac{\eta^2}{\bar a_r}\left[
\frac{\frac{14528 \nu }{315}+\frac{55424}{315}}{e_r^3}
+\frac{\left(\frac{16 \pi }{105}+\frac{1371 \pi ^3}{56}\right) \nu +\frac{160431\pi ^3}{1120}-\frac{98496 \pi }{175}}{e_r^4}
\right.\nonumber\\
&&
+\frac{\left(\frac{164096}{315}-\frac{2413568 \pi ^2}{11025}\right) \nu -\frac{7014 \zeta(3)}{5}+\frac{9149824 \pi ^2}{6615}+\frac{672316}{315}}{e_r^5}
\nonumber\\
&&\left.\left.
+\frac{\left(\frac{48 \pi }{35}+\frac{3968029 \pi ^3}{1680}-\frac{10735 \pi ^5}{56}\right) \nu -\frac{3050323 \pi ^5}{3584}+\frac{241678841 \pi^3}{24640}-\frac{2804288 \pi }{1575}}{e_r^6}
\right]
+O\left(\frac1{e_r^7}\right)
\right\}
\,.
\eea

It is also interesting to compute the change of the impact parameter $b= jMh/\sqrt{\gamma^2-1}$ due to the gravitational radiation damping, first evaluated in Ref. \cite{Junker:1992kle} at the 1PN approximation.
The expression for the radiated angular momentum appears misprinted in Ref. \cite{Junker:1992kle}, as already noticed in Refs. \cite{Bini:2021gat,Saketh:2021sri}.
This minor computational error then propagates in the subsequent calculations, so that our result \eqref{deltabinst} does not completely agree with Eq. (49) of Ref. \cite{Junker:1992kle} at the 1PN level.
We express it as a function of the gauge-invariant quantities $\bar E$ and $j$ to ease comparison.
We find
\bea
\label{deltabinst}
\left\langle\frac{db}{dt}\right\rangle^{\rm (hyp)}_{\rm inst}&=&\frac{\nu}{(2\bar E)^{3/2}}\left\{
A_1{\rm arccos}\left(-\frac1{\sqrt{1+2\bar Ej^2}}\right)+A_2\sqrt{2\bar E}
+\left[B_1{\rm arccos}\left(-\frac1{\sqrt{1+2\bar Ej^2}}\right)+B_2\frac{\sqrt{2\bar E}}{1+2\bar Ej^2}\right]\eta^2\right.\nonumber\\
&&\left.
+\left[C_1{\rm arccos}\left(-\frac1{\sqrt{1+2\bar Ej^2}}\right)+C_2\frac{\sqrt{2\bar E}}{(1+2\bar Ej^2)^2}\right]\eta^4
\right\}
\,,
\eea
with
\bea
A_1&=&\frac{170}{3 j^6}+\frac{124}{5 j^4} (2\bar E)-\frac{94}{15 j^2} (2\bar E)^2
\,,\nonumber\\
A_2&=&\frac{170}{3 j^5}+\frac{266}{45 j^3} (2\bar E)-\frac{16}{5 j} (2\bar E)^2
\,,\nonumber\\
B_1&=&\frac{\frac{13447}{20}-\frac{1127 \nu }{3}}{j^8}
+\frac{\frac{1177}{3}-\frac{4735 \nu }{12}}{j^6} (2\bar E)
+\frac{\frac{1833}{140}-\frac{521 \nu }{10}}{j^4} (2\bar E)^2
+\frac{\frac{329 \nu }{60}-\frac{2663}{210}}{j^2} (2\bar E)^3
\,,\nonumber\\
B_2&=&\frac{\frac{13447}{20}-\frac{1127 \nu }{3}}{j^7}
+\frac{\frac{25217}{30}-\frac{23221 \nu }{36}}{j^5} (2\bar E)
+\frac{\frac{1165511}{6300}-\frac{3976 \nu }{15}}{j^3} (2\bar E)^2
+\frac{\frac{341 \nu }{180}-\frac{56207}{3150}}{j} (2\bar E)^3
+j \left(\frac{4 \nu }{5}-\frac{204}{35}\right) (2\bar E)^4
\,,\nonumber\\
C_1&=&\frac{\frac{5481 \nu ^2}{4}-\frac{258051 \nu }{40}+\frac{5839651}{1008}}{j^{10}}
+\frac{\frac{57281 \nu ^2}{24}-\frac{9311813 \nu}{1440}+\frac{38419379}{12960}}{j^8} (2\bar E)
+\frac{\frac{200363 \nu ^2}{192}-\frac{126253 \nu }{84}+\frac{4357219}{12096}}{j^6} (2\bar E)^2\nonumber\\
&& 
+\frac{\frac{2165 \nu ^2}{32}-\frac{192709 \nu }{3360}+\frac{2330941}{15120}}{j^4} (2\bar E)^3
+\frac{-\frac{3713 \nu^2}{960}+\frac{2663 \nu }{336}+\frac{135421}{20160}}{j^2} (2\bar E)^4
\,,\nonumber\\
C_2&=&\frac{\frac{5481 \nu ^2}{4}-\frac{258051 \nu }{40}+\frac{5839651}{1008}}{j^9}
+\frac{\frac{112091 \nu ^2}{24}-\frac{24794873 \nu}{1440}+\frac{1144883303}{90720}}{j^7} (2\bar E)
+\frac{\frac{16566349 \nu ^2}{2880}-\frac{2377053491 \nu}{151200}+\frac{4567250873}{544320}}{j^5} (2\bar E)^2\nonumber\\
&& 
+\frac{\frac{8569237 \nu^2}{2880}-\frac{166914749 \nu }{30240}+\frac{36086381231}{19051200}}{j^3} (2\bar E)^3
+\frac{\frac{1493027 \nu ^2}{2880}-\frac{84736097 \nu}{151200}+\frac{7183074937}{19051200}}{j} (2\bar E)^4\nonumber\\
&& 
+j \left(-\frac{14461 \nu ^2}{2880}+\frac{105649 \nu}{8400}+\frac{55574633}{705600}\right) (2\bar E)^5
+j^3 \left(-\frac{2 \nu ^2}{5}-\frac{184}{315}\right) (2\bar E)^6
\,,
\eea
and
\bea
\left\langle\frac{db}{dt}\right\rangle^{\rm (hyp)}_{\rm hered}&=&\nu \left\{
\frac{-\frac{896}{45} p_\infty^3
+\left(\frac{2816 \nu }{63}-\frac{448}{45}\right) p_\infty^5}{j^3}\right.\nonumber\\
&&
+\pi\frac{\frac{21 \pi^2}{20} p_\infty^2+\left[\left(\frac{16}{105}+\frac{633 \pi ^2}{280}\right) \nu -\frac{3627 \pi ^2}{1120}-\frac{6336  }{175}\right] p_\infty^4}{j^4}\nonumber\\
&&
+\frac{\left(\frac{1792}{9}+\frac{23296 \pi ^2}{675}\right) p_\infty
+\left[\left(-\frac{43648}{63}-\frac{21921664 \pi ^2}{33075}\right) \nu -\frac{1218 \zeta (3)}{5}+\frac{3074432 \pi^2}{33075}+\frac{6196}{35}\right]p_\infty^2 }{j^5}\nonumber\\
&&\left.
+\pi\frac{-\frac{1063 \pi ^4}{64}+\frac{670 \pi ^2}{3}
+\left[\left(\frac{8 }{7}-\frac{33373 \pi ^2}{42}+\frac{24443 \pi ^4}{448}\right) \nu +\frac{286679 \pi ^4}{3584}-\frac{6381659 \pi^2}{7392}+\frac{209696 }{225}\right] p_\infty}{j^6}
+O\left(\frac1{j^6}\right)
\right\}
\,.\nonumber\\
\eea

\end{widetext}

\section{Concluding remarks}

We have analytically computed the orbital average of certain combinations of the source multipole moments, which are the building blocks entering the multipolar expansion of the radiated energy and angular momentum by gravitational waves.
Orbital averaging gives a gauge-invariant characterization of the radiation field associated with the source.
Therefore, such multipolar invariants can provide useful information to other approaches (EFT, amplitudes, etc.), which might benefit of all these explicit expressions to test their own (intermediate) results.

At the Newtonian level, the energy flux is expressed entirely in terms of the contraction of two quadrupole mass moments, each carrying three time derivatives. In the case of ellipticlike motion the orbital averaging then leads to the well known Peters-Mathews formula for the radiated energy, which is given by the quasicircular value times an eccentricity-dependent enhancement factor.
We have computed here the hyperboliclike counterparts of the Peters-Mathews enhancement factor and its PN corrections, which have been referred to as \lq\lq eccentricity enhancement factor functions" (EEFF), working at the 2.5PN level of accuracy.
To this end, we have also raised the accuracy of the angular momentum tail terms to the 1PN fractional order, also confirming the recent result by Cho \cite{Cho:2022pqy} for the energy tails by a fully analytic approach in the Fourier domain.
We have then used the complete 2.5PN-accurate averaged energy and angular momentum fluxes to compute the evolution of the orbital elements under gravitational radiation reaction.

Finally, we have expressed both sets of energy and angular momentum EEFF (most of them summarized in the various tables contained in the Appendices) in terms of the total conserved energy and angular momentum of the system, discussing then the correspondence between them by using general prescriptions (notably the works \cite{Kalin:2019rwq,Saketh:2021sri}) to convert information from the bound sector to the unbound one.

\section*{Acknowledgments}
The authors thank T. Damour for useful discussions at various stages during the development of the present project. DB thanks ICRANet for partial support and the IHES for warm hospitality in the completion phase of the work.
He also acknowledges sponsorship of the Italian Gruppo Nazionale per la Fisica Matematica (GNFM) of the Istituto Nazionale di Alta Matematica (INDAM).

\appendix

\section{Comparison between bound and unbound observables}

We list in the tables below the multipolar invariants $I_{l[r,r]}$ and $I^*_{z\,l[r,s]}$ entering the instantaneous contributions to the 2.5PN accurate expressions for the averaged energy and angular momentum fluxes along both ellipticlike and hyperboliclike orbits.


\begin{table*}  
\caption{\label{tab:tableIl_rr}  $I_{l[r,r]}$: ellipticlike vs hyperboliclike contributions. }
\begin{ruledtabular}
\begin{tabular}{ll}
$I_{3[4,4]}^{\rm (ell)}(\bar E, j)$  &$\pi\nu^2(1-4\nu)\left[A+B\eta^2+O(\eta^4)\right] $\\
$I_{3[4,4]}^{\rm (hyp)}(\bar E, j)$  &$\nu^2(1-4\nu)\left\{A{\rm arccos}\left(-\frac1{\sqrt{1+2\bar Ej^2}}\right)+A_2\sqrt{2\bar E}+\left[B{\rm arccos}\left(-\frac1{\sqrt{1+2\bar Ej^2}}\right)+B_2\frac{\sqrt{2\bar E}}{1+2\bar Ej^2}\right]\eta^2+O(\eta^4)\right\} $\\
$A$ &$-\frac{15273}{ 20 j^3} (-2\bar E)^3+\frac{347589}{ 20 j^5} (-2\bar E)^2-\frac{218715}{ 4 j^7} (-2\bar E)+\frac{165375}{ 4 j^9}$\\
$B$ &$\frac{(\frac{116253}{80}\nu-\frac{6165}{8})}{ j^3} (-2\bar E)^4
+\frac{(-\frac{1320057}{20}\nu+\frac{471747}{10})}{ j^5} (-2\bar E)^3
+\frac{(\frac{6900801}{16}\nu-\frac{3591513}{16})}{ j^7} (-2\bar E)^2
+\frac{(-\frac{6847799}{8}\nu+\frac{6544643}{40})}{ j^9} (-2\bar E)$\\
&$
+\frac{(512820\nu+\frac{1476981}{16})}{ j^{11}}$\\
$A_2$&$ \frac{148439}{ 20 j^4} (2\bar E)^2+\frac{81795}{ 2 j^6} (2\bar E)+\frac{165375}{ 4 j^8}$\\
$B_2$ &$ \frac{(\frac{1646499}{80}\nu-\frac{20399621}{1400})}{ j^2} (2\bar E)^4
+\frac{(-\frac{356434399}{2100}+\frac{32245229}{120}\nu)}{ j^4} (2\bar E)^3
+\frac{(-\frac{83005229}{240}+\frac{44811551}{48}\nu)}{ j^6} (2\bar E)^2
+\frac{(-\frac{510376}{5}+\frac{9582839}{8}\nu)}{ j^8} (2\bar E)$\\
&$
+\frac{(512820\nu+\frac{1476981}{16})}{ j^{10}}$\\
\hline
$J_{2[3,3]}^{\rm (ell)}(\bar E, j)$  &$\pi\nu^2(1-4\nu)\left[A+B\eta^2+O(\eta^4)\right] $\\
$J_{2[3,3]}^{\rm (hyp)}(\bar E, j)$  &$\nu^2(1-4\nu)\left\{A{\rm arccos}\left(-\frac1{\sqrt{1+2\bar Ej^2}}\right)+A_2\sqrt{2\bar E}+\left[B{\rm arccos}\left(-\frac1{\sqrt{1+2\bar Ej^2}}\right)+B_2\frac{\sqrt{2\bar E}}{1+2\bar Ej^2}\right]\eta^2+O(\eta^4)\right\} $\\
$A$&$ -\frac{9}{ 16 j^3} (-2\bar E)^3+\frac{165}{ 16 j^5} (-2\bar E)^2-\frac{455}{ 16 j^7} (-2\bar E)+\frac{315}{ 16 j^9}$\\
$B$&$ \frac{(\frac{9}{56}\nu-\frac{4113}{1792})}{ j^3} (-2\bar E)^4
+\frac{(-\frac{615}{56}\nu+\frac{30145}{448})}{ j^5} (-2\bar E)^3
+\frac{(\frac{5055}{64}\nu-\frac{30535}{128})}{ j^7} (-2\bar E)^2
+\frac{(-\frac{5229}{32}\nu+\frac{10773}{64})}{ j^9} (-2\bar E)
+\frac{(\frac{6435}{64}\nu+\frac{10725}{256})}{j^{11}}$\\
$A_2$&$ \frac{229}{ 48 j^4} (2\bar E)^2+\frac{175}{ 8 j^6} (2\bar E)+\frac{315}{ 16/j^8}$\\
$B_2$&$ \frac{(\frac{867}{280}\nu-\frac{706817}{26880})}{j^2} (2\bar E)^4
+\frac{(-\frac{2707981}{13440}+\frac{13327}{280}\nu)}{j^4} (2\bar E)^3
+\frac{(-\frac{22807}{64}+\frac{11169}{64}\nu)}{j^6} (2\bar E)^2
+\frac{(-\frac{17971}{128}+\frac{3687}{16}\nu)}{j^8} (2\bar E)
+\frac{(\frac{6435}{64}\nu+\frac{10725}{256})}{j^{10}}$\\
\hline
$I_{4[5,5]}^{\rm (ell)}(\bar E, j)$  &$\pi\nu^2(1-3\nu)^2\left[A+O(\eta^2)\right] $\\
$I_{4[5,5]}^{\rm (hyp)}(\bar E, j)$  &$\nu^2(1-3\nu)^2\left\{\left[A{\rm arccos}\left(-\frac1{\sqrt{1+2\bar Ej^2}}\right)+A_2\sqrt{2\bar E}\right]+O(\eta^2)\right\} $\\
$A$&$ \frac{2347659}{ 70 j^3} (-2\bar E)^4-\frac{50808042}{ 35 j^5} (-2\bar E)^3+\frac{62748897}{ 7 j^7} (-2\bar E)^2-\frac{16795030}{ j^9} (-2\bar E)+\frac{19022625}{ 2 j^{11}}$\\
$A_2$&$ \frac{4638627}{ 10 j^4} (2\bar E)^3
+\frac{221257987}{ 42 j^6 }(2\bar E)^2
+\frac{27249185}{ 2 j^8} (2\bar E)
+\frac{19022625}{ 2 j^{10}}$\\
\hline
$J_{3[4,4]}^{\rm (ell)}(\bar E, j)$  &$\pi\nu^2(1-3\nu)^2\left[A+O(\eta^2)\right] $\\
$J_{3[4,4]}^{\rm (hyp)}(\bar E, j)$  &$\nu^2(1-3\nu)^2\left\{\left[A{\rm arccos}\left(-\frac1{\sqrt{1+2\bar Ej^2}}\right)+A_2\sqrt{2\bar E}\right]+O(\eta^2)\right\} $\\
$A$&$ \frac{993}{ 40 j^3} (-2\bar E)^4-\frac{5273}{ 6 j^5} (-2\bar E)^3+\frac{57449}{ 12 j^7} (-2\bar E)^2-\frac{82467}{10 j^9} (-2\bar E)+\frac{35189}{8 j^{11}}$\\
$A_2$&$ \frac{547187}{ 1800 j^4} (2\bar E)^3+\frac{350189}{ 120 j^6} (2\bar E)^2+\frac{813659}{120 j^8} (2\bar E)+\frac{35189}{ 8 j^{10}}$\\
\end{tabular}
\end{ruledtabular}
\end{table*}


\begin{table*}  
\caption{\label{tab:tableIstarl_rs}  $I^*_{z\,l[r,s]}$: ellipticlike vs hyperboliclike contributions. }
\begin{ruledtabular}
\begin{tabular}{ll}
$I_{z\,3[3,4]}^{*\,\rm (ell)}(\bar E, j)$  & $\pi\nu^2(1-4\nu)\left[A+B\eta^2+O(\eta^4)\right] $\\
$I_{z\,3[3,4]}^{*\,\rm (hyp)}(\bar E, j)$  &$\nu^2(1-4\nu)\left\{A{\rm arccos}\left(-\frac1{\sqrt{1+2\bar Ej^2}}\right)+A_2\sqrt{2\bar E}+\left[B{\rm arccos}\left(-\frac1{\sqrt{1+2\bar Ej^2}}\right)+B_2\frac{\sqrt{2\bar E}}{1+2\bar Ej^2}\right]\eta^2+O(\eta^4)\right\} $\\
$A$ &$\frac{5301}{ 10 j^2} (2\bar E)^2+\frac{4389}{j^4} (2\bar E)+\frac{9905}{ 2 j^6} $\\
$B$ &$\frac{(\frac{13191}{20}\nu-\frac{1641}{20})}{j^2} (2\bar E)^3+\frac{(\frac{341741}{20}\nu-\frac{44267}{4})}{j^4} (2\bar E)^2
+\frac{(\frac{121429}{2}\nu-\frac{52451}{2})}{j^6} (2\bar E)
+\frac{(\frac{151760}{3}\nu-\frac{10129}{3})}{j^8} $\\
$A_2$&$\frac{288}{ 5 j} (2\bar E)^2+\frac{16429}{ 6 j^3} (2\bar E)+\frac{9905}{ 2 j^5} $\\
$B_2$ &$ (\frac{84}{5}\nu+\frac{372}{5}) j (2\bar E)^4+\frac{(\frac{410779}{60}\nu-\frac{49873}{12})}{j} (2\bar E)^3
+\frac{(\frac{9147329}{180}\nu-\frac{5058043}{180})}{j^3} (2\bar E)^2+\frac{(\frac{1699901}{18}\nu-\frac{512575}{18})}{j^5} (2\bar E)$\\
&$
+\frac{(\frac{151760}{3}\nu-\frac{10129}{3})}{j^7}$\\
\hline
$J_{z\,2[2,3]}^{*\,\rm (ell)}(\bar E, j)$  & $\pi\nu^2(1-4\nu)\left[A+B\eta^2+O(\eta^4)\right] $\\
$J_{z\,2[2,3]}^{*\,\rm (hyp)}(\bar E, j)$  &$\nu^2(1-4\nu)\left\{A{\rm arccos}\left(-\frac1{\sqrt{1+2\bar Ej^2}}\right)+A_2\sqrt{2\bar E}+\left[B{\rm arccos}\left(-\frac1{\sqrt{1+2\bar Ej^2}}\right)+B_2\frac{\sqrt{2\bar E}}{1+2\bar Ej^2}\right]\eta^2+O(\eta^4)\right\} $\\
$A$ &$\frac{3}{16j^2} (2\bar E)^2+\frac{15}{ 8 j^4} (2\bar E)+\frac{35}{ 16j^6}$\\
$B$ &$\frac{\frac{9}{224}\nu-\frac{345}{448}}{ j^2} (2\bar E)^3+\frac{\frac{435}{224}\nu-\frac{4425}{448}}{ j^4} (2\bar E)^2+\frac{\frac{35}{4}\nu-\frac{595}{64}}{ j^6} (2\bar E)+\frac{\frac{135}{16}\nu+\frac{549}{64}}{ j^8}$\\
$A_2$&$\frac{55}{48j^3} (2\bar E)+\frac{35}{16j^5}$\\
$B_2$ &$ \frac{\frac{437}{672}\nu-\frac{37801}{6720}}{j} (2\bar E)^3+\frac{\frac{4469}{672}\nu-\frac{115711}{6720}}{j^3} (2\bar E)^2+\frac{\frac{115}{8}\nu-\frac{229}{64}}{j^5} (2\bar E)+\frac{\frac{135}{16}\nu+\frac{549}{64}}{j^7}$\\
\hline
$I_{z\,4[4,5]}^{*\,\rm (ell)}(\bar E, j)$  & $\pi\nu^2(1-3\nu)^2\left[A+O(\eta^2)\right] $\\
$I_{z\,4[4,5]}^{*\,\rm (hyp)}(\bar E, j)$  &$\nu^2(1-3\nu)^2\left\{A{\rm arccos}\left(-\frac1{\sqrt{1+2\bar Ej^2}}\right)+A_2\sqrt{2\bar E}+O(\eta^2)\right\} $\\
$A$ &$\frac{107829}{ 7 j^2}(2\bar E)^3+\frac{2287227}{ 7 j^4}(2\bar E)^2+\frac{1033545}{j^6}(2\bar E)+\frac{787775}{j^8}$\\
$A_2$&$\frac{4608}{ 7 j} (2\bar E)^3+\frac{978507}{ 7 j^3}(2\bar E)^2+\frac{2312860}{3j^5}(2\bar E)+\frac{787775}{j^7}$\\
\hline
$J_{z\,3[3,4]}^{*\,\rm (ell)}(\bar E, j)$  & $\pi\nu^2(1-3\nu)^2\left[A+O(\eta^2)\right] $\\
$J_{z\,3[3,4]}^{*\,\rm (hyp)}(\bar E, j)$  &$\nu^2(1-3\nu)^2\left\{A{\rm arccos}\left(-\frac1{\sqrt{1+2\bar Ej^2}}\right)+A_2\sqrt{2\bar E}+O(\eta^2)\right\} $\\
$A$ &$\frac{23}{ 3 j^2} (2\bar E)^3+\frac{505}{ 3 j^4} (2\bar E)^2+\frac{4655}{ 9 j^6} (2\bar E)+\frac{385}{ j^8}$\\
$A_2$&$\frac{1969}{ 27 j^3} (2\bar E)^2+\frac{3500}{ 9 j^5} (2\bar E)+\frac{385}{ j^7}$\\
\end{tabular}
\end{ruledtabular}
\end{table*}

\section{Radiated energy and angular momentum along hyperboliclike orbits at the 2.5PN level}

The instantaneous contributions to the radiated energy and angular momentum through the 2PN order are given by the following exact expressions

\begin{widetext}

\bea
(\Delta E)_{\rm inst}&=&\nu^2\left\{
A_1{\rm arccos}\left(-\frac1{\sqrt{1+2\bar Ej^2}}\right)+A_2\sqrt{2\bar E}
+\left[B_1{\rm arccos}\left(-\frac1{\sqrt{1+2\bar Ej^2}}\right)+B_2\frac{\sqrt{2\bar E}}{1+2\bar Ej^2}\right]\eta^2\right.\nonumber\\
&&\left.
+\left[C_1{\rm arccos}\left(-\frac1{\sqrt{1+2\bar Ej^2}}\right)+C_2\frac{\sqrt{2\bar E}}{(1+2\bar Ej^2)^2}\right]\eta^4
\right\}
\,,
\eea
with
\bea
A_1&=&\frac{170}{3 j^7}+\frac{244 }{5 j^5}(2\bar E)+\frac{74 }{15 j^3}(2\bar E)^2
\,,\nonumber\\
A_2&=&\frac{170}{3 j^6}+\frac{1346}{45 j^4} (2\bar E)
\,,\nonumber\\
B_1&=&\frac{\frac{13447}{20}-\frac{1127 \nu }{3}}{j^9}+\frac{\frac{2259}{4}-530 \nu }{j^7}(2\bar E)+\frac{\frac{15539}{140}-\frac{872 \nu }{5}}{j^5}(2\bar E)^2+\frac{\frac{2393}{420}-\frac{37 \nu }{5}}{j^3}(2\bar E)^3
\,,\nonumber\\
B_2&=& \frac{\frac{13447}{20}-\frac{1127 \nu}{3}}{j^8}+\frac{\frac{60779}{60}-\frac{7024 \nu }{9}}{j^6}(2\bar E)+\frac{\frac{835477}{2100}-\frac{21494 \nu }{45}}{j^4}(2\bar E)^2+\frac{\frac{29969}{700}-\frac{1117 \nu }{15}}{j^2}(2\bar E)^3
\,,\nonumber\\
C_1&=&\frac{\frac{5481 \nu ^2}{4}-\frac{258051 \nu}{40}+\frac{5839651}{1008}}{j^{11}}+\frac{\frac{5733 \nu ^2}{2}-\frac{2838577\nu }{360}+\frac{11947909}{3240} }{j^9}(2\bar E)+\frac{\frac{7075 \nu^2}{4}-\frac{862691 \nu }{336}+\frac{736055}{3024}}{j^7}(2\bar E)^2\nonumber\\
&& 
+\frac{\frac{3201 \nu ^2}{10}-\frac{109657 \nu}{420}+\frac{2995}{126} }{j^5}(2\bar E)^3+\frac{\frac{185 \nu^2}{24}-\frac{2393 \nu }{336}+\frac{149}{18}}{j^3}(2\bar E)^4
\,,\nonumber\\
C_2&=&\frac{\frac{5481 \nu ^2}{4}-\frac{258051 \nu}{40}+\frac{5839651}{1008}}{j^{10}}+\frac{\frac{20601 \nu^2}{4}-\frac{3354671 \nu }{180}+\frac{605244551}{45360}}{j^8}(2\bar E)+\frac{\frac{145541 \nu ^2}{20}-\frac{1447722833 \nu}{75600}+\frac{1289930977}{136080} }{j^6}(2\bar E)^2\nonumber\\
&& 
+\frac{\frac{56309 \nu^2}{12}-\frac{17828819 \nu }{2160}+\frac{9310241671}{4762800}}{j^4}(2\bar E)^3+\frac{\frac{156113 \nu ^2}{120}-\frac{103927661 \nu}{75600}+\frac{13061807}{297675} }{j^2}(2\bar E)^4\nonumber\\
&& 
+\left(\frac{4231 \nu^2}{40}-\frac{680471 \nu }{8400}+\frac{509759}{14700}\right) (2\bar E)^5
\,,
\eea
and
\bea
(\Delta J_z)_{\rm inst}&=&\nu^2\left\{
A_1{\rm arccos}\left(-\frac1{\sqrt{1+2\bar Ej^2}}\right)+A_2\sqrt{2\bar E}
+\left[B_1{\rm arccos}\left(-\frac1{\sqrt{1+2\bar Ej^2}}\right)+B_2\frac{\sqrt{2\bar E}}{1+2\bar Ej^2}\right]\eta^2\right.\nonumber\\
&&\left.
+\left[C_1{\rm arccos}\left(-\frac1{\sqrt{1+2\bar Ej^2}}\right)+C_2\frac{\sqrt{2\bar E}}{(1+2\bar Ej^2)^2}\right]\eta^4
\right\}
\,,
\eea
with
\bea
A_1&=&\frac{56}{ 5 j^2}(2\bar E)+\frac{24}{j^4}
\,,\nonumber\\
A_2&=&\frac{16}{ 5 j}(2\bar E)+\frac{24}{ j^3}
\,,\nonumber\\
B_1&=&\frac{-\frac{284}{15}\nu+\frac{4283}{210}}{j^2} (2\bar E)^2+\frac{107-\frac{748}{5}\nu}{j^4} (2\bar E)+\frac{\frac{359}{2}-\frac{470}{3}\nu}{j^6}
\,,\nonumber\\
B_2&=& \left(\frac{218}{35}-2\nu\right) j (2\bar E)^3+\frac{-\frac{880}{9} \nu+\frac{14237}{210}}{ j} (2\bar E)^2+\frac{\frac{680}{3}-\frac{11432}{45}\nu}{j^3}(2\bar E)+\frac{\frac{359}{2}-\frac{470}{3}\nu}{j^5}
\,,\nonumber\\
C_1&=&\frac{\frac{1327}{315}-\frac{3653}{140}\nu+\frac{337}{15}\nu^2}{j^2} (2\bar E)^3
+\frac{-\frac{131321}{420}\nu+\frac{1903}{5}\nu^2-\frac{20249}{189}}{j^4} (2\bar E)^2\nonumber\\
&& 
+\frac{-\frac{8497}{6}\nu-\frac{2999}{108}+\frac{2968}{3}\nu^2}{j^6} (2\bar E)
+\frac{\frac{1307683}{1620}+\frac{1862}{3}\nu^2-\frac{77287}{45}\nu}{j^8}
\,,\nonumber\\
C_2&=&\left(\frac{649}{504}-\frac{327}{140}\nu+\frac{51}{40}\nu^2\right) j^3 (2\bar E)^5
+\left(-\frac{406927}{12600}-\frac{471853}{3150}\nu+\frac{65099}{360}\nu^2\right) j (2\bar E)^4\nonumber\\
&&
+\frac{-\frac{10659386}{42525}+\frac{51172}{45}\nu^2-\frac{11072003}{9450}\nu}{j} (2\bar E)^3
+\frac{\frac{47133773}{170100}+\frac{106223}{45}\nu^2-\frac{67830983}{18900}\nu}{j^3} (2\bar E)^2\nonumber\\
&&
+\frac{\frac{320173}{243}+\frac{18214}{9}\nu^2-\frac{231047}{54}\nu}{j^5} (2\bar E)
+\frac{\frac{1307683}{1620}+\frac{1862}{3}\nu^2-\frac{77287}{45}\nu}{j^7}
\,,
\eea

\end{widetext}
respectively. 
The contribution of tails is 1.5PN and 2.5PN order (Eq. \eqref{1PNtails}), known as a large-$j$ expansion only up to the N$^3$LO.
In the case of angular momentum, one has to add also the 2.5PN memory term
\bea
\label{j_mem}
(\Delta J_z)_{\rm mem}&=&-\frac{GM^2}{c}\nu^3\left[ \frac{16}{105}\frac{p_\infty^5\pi}{j^4} + \frac{128}{63}\frac{p_\infty^4}{j^5}\right.\nonumber\\
&&\left.
 + \frac{8}{7}\frac{p_\infty^3\pi}{j^6}+O\left(\frac{1}{j^7}\right)\right]\,,
\eea
computed in Ref. \cite{Bini:2021qvf} in the same limit.

The radiated energy and angular momentum along hyperboliclike orbits admit the following double PM-PN expansion \cite{Bini:2021gat}
\bea
\frac{E^{\rm rad}}{M} &=&   \nu^2  \left[ \frac{{E}_{3}(p_\infty)}{j^3}+ \frac{{ E}_{4}(p_\infty)}{j^4}+  \cdots\right]  \,, \nonumber\\
\frac{ J_i^{\rm rad}}{J} &=&  \nu \left[ \frac{{J}_{i\,2}(p_\infty)}{j^2}+ \frac{{J}_{i\,3}(p_\infty)}{j^3} + \cdots\right]\,, \eea
where $p_{\infty}$ and $j$ are convenient dimensionless variables representing the incoming (at $t\to -\infty$, i.e., before the starting of the scattering process) energy and angular momentum, respectively, while the subscripts $n=2,3,4,\ldots$ in the coefficients $E_n$ and $J_{i\,n}$ label the $n$PM order, i.e., $O(G^n)$.
The first few PM expansion coefficients of the energy loss have the following structure
\bea
{ E}_{3}(p_\infty)&\sim& \underbrace{p_\infty^4}_{\rm N}+ \underbrace{p_\infty^6}_{\rm 1PN}+ \underbrace{p_\infty^8}_{\rm 2PN}+\ldots
\,,\nonumber\\
{ E}_{4}(p_\infty)&\sim& \underbrace{p_\infty^3}_{\rm N}+ \underbrace{p_\infty^5}_{\rm 1PN}+ \underbrace{p_\infty^6}_{\rm N\, tail}+ \underbrace{p_\infty^7}_{\rm 2PN}+\underbrace{p_\infty^8}_{\rm 1PN\, tail}\ldots
\,,\nonumber\\
{ E}_{5}(p_\infty)&\sim& \underbrace{p_\infty^2}_{\rm N}+ \underbrace{p_\infty^4}_{\rm 1PN}+ \underbrace{p_\infty^5}_{\rm N\, tail}+\underbrace{p_\infty^6}_{\rm 2PN}\nonumber\\
&+ & \underbrace{p_\infty^7}_{\rm 1PN\, tail}+\underbrace{p_\infty^8}_{\rm 3PN + tail(tail)+(tail)^2}+\ldots\,,
\nonumber\\
\eea
where the expansion in powers of $p_\infty$ corresponds to the usual PN expansion.
Similar expressions hold for the angular momentum loss (with only nonvanishing component along the $z$-axis with the present choice of coordinates) 
\bea
{J}_{z\,2}(p_\infty)&\sim& \underbrace{p_\infty^3}_{\rm N}+ \underbrace{p_\infty^5}_{\rm 1PN}+ \underbrace{p_\infty^7}_{\rm 2PN}+\ldots
\,,\nonumber\\
{J}_{z\,3}(p_\infty)&\sim& \underbrace{p_\infty^2}_{\rm N}+ \underbrace{p_\infty^4}_{\rm 1PN}+ \underbrace{p_\infty^6}_{\rm 2PN}+\ldots
\,,\nonumber\\
{J}_{z\,4}(p_\infty)&\sim& \underbrace{p_\infty}_{\rm N}+ \underbrace{p_\infty^3}_{\rm 1PN}+\underbrace{p_\infty^4}_{\rm N\, tail}+ \underbrace{p_\infty^5}_{\rm 2PN}+\underbrace{p_\infty^6}_{\rm 1PN\, tail}+\ldots
\,. \nonumber\\
\eea
Up to the order $O(G^3)$, i.e., at the 3PM level of accuracy, all these coefficients are analytically known (in PM sense, i.e., resumming all the PN orders) thanks to the recent results achieved in the framework of effective field theory (EFT) \cite{Herrmann:2021tct,Manohar:2022dea}, pioneered by Damour \cite{Damour:2020tta}. 
For example, Ref. \cite{Manohar:2022dea} has obtained a closed form expression for ${J}_{z\,3}$, whose high-PN expansion reads
\bea
\frac{{J}_{z\,3}(p_\infty)}{\pi}&=&
\frac{28}{5} p_\infty^2
+\left(\frac{739}{84} -\frac{163}{15}\nu \right)p_\infty^4\nonumber\\
&+&\left( -\frac{5777}{2520} -\frac{5339}{420}\nu +\frac{50}{3}\nu^2\right)p_\infty^6\nonumber\\
&+&
\left( \frac{115769}{126720} +\frac{1469}{504}\nu +\frac{9235}{672}\nu^2 -\frac{553}{24} \nu^3\right)p_\infty^8\nonumber\\
&+&
\left(-\frac{16548173}{23063040} -\frac{292139}{221760}\nu-\frac{5183}{2016}\nu^2\right.\nonumber\\
&&\left. -\frac{563}{48} \nu^3+\frac{959}{32} \nu^4\right)p_\infty^{10}
+O(p_\infty^{12})\,.
\eea
The first three terms only of the above expression were already known (see Table IX of Ref. \cite{Bini:2021gat}, where the various coefficients are listed at the fractional 2PN accuracy through to the seventh order in $G$, i.e., up to $n=7$).

We are then able to improve the PN accuracy of these coefficients by including the contribution of the 1PN tails.
We provide in Table \ref{tab:newcoeffsEnJn} below the newly computed 2.5PN terms only.


\begin{table*}  
\caption{\label{tab:newcoeffsEnJn} 2.5PN terms improving the PN expansion of the coefficients $E_n$ and $J_n$ given in Table IX of Ref. \cite{Bini:2021gat}.
}
\begin{ruledtabular}
\begin{tabular}{ll}
$E_{4}^{\rm 2.5PN}$&$\left(\frac{1216}{105}-\frac{2848}{15}\nu\right)p_\infty^8 $\\
$E_{5}^{\rm 2.5PN}$&$\pi\left( -\frac{15291 \pi ^2 \nu }{280}-\frac{24993 \pi^2}{1120}+\frac{9216}{35}\right)p_\infty^7$\\
$E_{6}^{\rm 2.5PN}$&$ \left(-\frac{3014912 \pi ^2 \nu }{4725}-\frac{52928 \nu }{45}+\frac{2898 \zeta (3)}{5}+\frac{1024 \pi ^2}{135}+\frac{56708}{105}\right)p_\infty^6$\\
$E_{7}^{\rm 2.5PN}$&$ \pi\left(\frac{30285 \pi ^4 \nu }{112}-\frac{23514 \pi ^2 \nu }{7}+\frac{689985 \pi ^4}{3584}-\frac{13138915 \pi ^2}{7392}+\frac{210176}{225}\right)p_\infty^5$\\
\hline
$J_{4}^{\rm 2.5PN}$&$ \left(\frac{1184}{21}-\frac{17600 \nu }{63}\right)p_\infty^6$\\
$J_{5}^{\rm 2.5PN}$&$ \pi\left(-\frac{2232 \pi ^2 \nu }{35}-\frac{16 \nu }{105}-\frac{1305 \pi ^2}{112}+\frac{7488}{25}\right)p_\infty^5$\\
$J_{6}^{\rm 2.5PN}$&$ \left(-\frac{156416 \pi ^2 \nu }{6615}-\frac{10240 \nu }{21}+\frac{4116 \zeta (3)}{5}-\frac{130688 \pi ^2}{6615}+\frac{147064}{315}\right)p_\infty^4$\\
$J_{7}^{\rm 2.5PN}$&$ \pi\left(\frac{102619 \pi ^4 \nu }{448}-\frac{57037 \pi ^2 \nu }{21}-\frac{8 \nu }{7}+\frac{163083 \pi ^4}{1792}-\frac{18227 \pi ^2}{28}+\frac{32}{15}\right)p_\infty^3$\\
\end{tabular}
\end{ruledtabular}
\end{table*}

\section {High-order eccentricity expansions of Newtonian hereditary EEFF along ellipticlike orbits}

Ref. \cite{Arun:2007rg} has shown how the hereditary part of the gravitational wave flux can be written in terms of various EEFF, as we have discussed above. These functions, denoted by Greek letters, fail in general any analytic closed form representation, and are obtained either numerically or in the form of power series of (any type) small eccentricity. At the Newtonian level it is not difficult to provide eccentricity expansions up to very high orders. 
We show in Tables \ref{tab:table_phi_beta_gamma}--\ref{tab:table_chi_et} the series expansion of the functions $\varphi(e_t)$, $\beta(e_t)$, $\gamma(e_t)$ and $\chi(e_t)$ (defined in Eqs. 5.4, 5.6a, 5.6b, 5.12 of Ref. \cite{Arun:2007rg}, respectively) in the small eccentricity truncated at $O(e_t^{30})$.

We notice that recently a considerable effort has been given to the determination of the energy and angular momentum losses by nonspinning extreme-mass-ratio inspirals along eccentric orbits within the GSF framework through a very high PN level and in a small-eccentricity expansion up to the order $O(e_t^{20})$ \cite{Munna:2020juq,Munna:2020iju}.


\begin{table*}  
\caption{\label{tab:table_phi_beta_gamma} EEFF $\varphi(e_t)$, $\beta(e_t)$, $\gamma(e_t)$  evaluated along ellipticlike orbits and expanded in series of small eccentricity, truncated at $O(e_t^{30})$. }
\begin{ruledtabular}
\begin{tabular}{ll}
$\varphi(e_t)$  & $1 + \frac{2335}{192} e_t^2 + \frac{42955}{768} e_t^4 + \frac{6204647}{36864} e_t^6 + \frac{352891481}{884736} e_t^8 + \frac{286907786543}{353894400} e_t^{10}+ \frac{6287456255443}{4246732800} e_t^{12} + \frac{5545903772613817}{2219625676800} e_t^{14}$   \\
& $ + \frac{422825073954708079}{106542032486400} e_t^{16}+ \frac{1659160118498286776339}{276156948204748800} e_t^{18}
+\frac{724723372042305454448081}{82847084461424640000}e_t^{20} + \frac{395596016464873396514172181}{32078391103463620608000}e_t^{22}$\\
&$ + \frac{13025923416821727455962629769}{769881386483126894592000}e_t^{24} + \frac{708245346456804863545731982149679}{31226389035755626844651520000}e_t^{26} + \frac{182399987908827322948150653676542901}{6120372251008102861551697920000}e_t^{28}
+O(e_t^{30})$  \\
\hline
$\beta(e_t)$ &$1 + \frac{303396}{16403} e_t^2 + \frac{122593417}{1049792} e_t^4 + \frac{4245487415}{9448128} e_t^6 + \frac{174617565355}{134373376} e_t^8+ \frac{47178621858259}{15117004800} e_t^{10} + \frac{25457292864087389}{3869953228800} e_t^{12}$   \\
&$ + \frac{1022978928834610541}{81269017804800} e_t^{14}+ \frac{6511979627941130726839}{291268159812403200} e_t^{16} + \frac{66228086083334497666747}{1769454070860349440} e_t^{18}
+ \frac{2254162867636735292910226543}{37748353511687454720000} e_t^{20} $\\
&$
+ \frac{1254430161020403927369484292543}{13702652324742546063360000} e_t^{22} + \frac{2142152999007599476852619853176963}{15785455478103413064990720000} e_t^{24} + \frac{1931310773208071115672891358826803}{9880525836294358548086784000}e_t^{26}$\\
&$ + \frac{1838162195484460130198576684064789412889}{6692831068885727415868833398784000}e_t^{28}
+O(e_t^{30})$   \\
\hline
$\gamma(e_t)$ & 
$ 1 + 30 e_t^2 + \frac{7321}{32} e_t^4 + \frac{279485}{288} e_t^6 + \frac{73162261}{24576} e_t^8 + \frac{3426964009}{460800} e_t^{10}+ \frac{105664935873}{6553600} e_t^{12} +\frac{273205556708681}{8670412800} e_t^{14}$   \\
&$  + \frac{144317835494605679}{2536715059200} e_t^{16}+ \frac{41638723329613052819}{431495231569920} e_t^{18}
+ \frac{179061139282005235389013}{1150653950853120000} e_t^{20} + \frac{20107796748815662814890339}{83537476831936512000} e_t^{22}$\\
&$ + \frac{345972689370635704746575577163}{962351733103908618240000} e_t^{24} 
+ \frac{706365395346807528358979578763}{1355312024121337970688000} e_t^{26} + \frac{187796011016826049541560927726232737}{255015510458670952564654080000} e_t^{28}
+O(e_t^{30})$  \\
\end{tabular}
\end{ruledtabular}
\end{table*}


\begin{table*}  
\caption{\label{tab:table_chi_et} EEFF $\chi(e_t)$ evaluated along ellipticlike orbits and expanded in series of small eccentricity, truncated at $O(e_t^{30})$.}
\begin{ruledtabular}
\begin{tabular}{ll}
$\chi(e_t)$ &$e_t{}^2
   \left(\frac{6561 \log (3)}{256}-\frac{77 \log (2)}{3}\right)
+e_t{}^4 \left(\frac{72533 \log
   (2)}{192}-\frac{124659 \log (3)}{1024}\right)
+e_t{}^6 \left(-\frac{1215829 \log
   (2)}{384}+\frac{3776949 \log (3)}{16384}+\frac{244140625 \log
   (5)}{147456}\right)$   \\
& $+e_t{}^8 \left(\frac{36266899 \log
   (2)}{3072}+\frac{1058532057 \log (3)}{131072}-\frac{10498046875 \log
   (5)}{1179648}\right)$   \\
& $
+e_t{}^{10} \left(-\frac{81758451257 \log
   (2)}{1382400}-\frac{5036206482351 \log (3)}{104857600}+\frac{2342041015625
   \log (5)}{113246208}+\frac{33232930569601 \log
   (7)}{943718400}\right)$   \\
& $+e_t{}^{12} \left(\frac{284916786743 \log
   (2)}{552960}+\frac{10365418177083 \log (3)}{83886080}-\frac{12554931640625
   \log (5)}{452984832}-\frac{33232930569601 \log
   (7)}{150994944}\right)$  \\
& $+e_t{}^{14} \left(-\frac{780705975924439
   \log (2)}{270950400}+\frac{119249295010191459 \log
   (3)}{164416716800}+\frac{19376068115234375 \log
   (5)}{799065243648}+\frac{202022984932604479 \log
   (7)}{326149079040}\right)$   \\
&$+e_t{}^{16}
   \left(\frac{382381662329585969 \log
   (2)}{39016857600}-\frac{16759832223639183561 \log
   (3)}{2630667468800}+\frac{18559246063232421875 \log
   (5)}{12785043898368}-\frac{27385031476060020833 \log
   (7)}{26091926323200}\right)$   \\
&$+e_t{}^{18} \left(-\frac{8052417982167544919 \log
   (2)}{300987187200}+\frac{909790206944625389457 \log
   (3)}{42090679500800}-\frac{330999693393707275390625 \log
   (5)}{29456741141839872}\right. $   \\
&$\left.+\frac{54564106085497121807827 \log
   (7)}{45086848686489600}+\frac{9849732675807611094711841 \log
   (11)}{2209255585637990400}\right)$\\  
&$+e_t^{20}\left(\frac{ 108338510145165255268981\ln(2)}{1264146186240000} - \frac{132376292038811899361331\ln(3)}{4209067950080000} + \frac{42528669410610198974609375\ln(5)}{1060442681106235392}\right.$\\
&$\left. - \frac{1531019455667272209191567\ln(7)}{1502894956216320000} - \frac{1605506426156640608438030083\ln(11)}{44185111712759808000}\right) $\\ 
&$+ e_t^{22}\left(-\frac{33827646161396251701627383\ln(2)}{101974459023360000} - \frac{633487618122696466753659561\ln(3)}{13038008882167808000} - \frac{16606595965492725372314453125\ln(5)}{186637911874697428992}\right.$\\
 &$+ \frac{  780894869749989113985610451\ln(7)}{1190292805323325440000} + \frac{1471540212032981289938854333559\ln(11)}{10604426811062353920000}
\left. + \frac{15502932802662396215269535105521\ln(13)}{42771188137951494144000}\right) $\\
&$+ e_t^{24} \left(\frac{8525792178752688986036947387\ln(2)}{7342161049681920000} + \frac{202485382033957177127957928693\ln(3)}{521520355286712320000} + \frac{2275226862233102321624755859375\ln(5)}{16424136244973373751296}\right.$\\
&$\left. + \frac{6152817171731901841962529296881\ln(7)}{62847460121071583232000} - \frac{28154308435072986823735716213103\ln(11)}{84835414488498831360000} - \frac{1131714094594354923714676062703033\ln(13)}{3421695051036119531520000}\right)$\\ 
&$+ e_t^{26}\left(-\frac{25689305949089519002862266157299\ln(2)}{7444951304377466880000} - \frac{42965619211204379451125391134259\ln(3)}{50363965739116789760000} + \frac{9702695818384550511837005615234375\ln(5)}{99924444914418005902884864}\right. $\\
&$ - \frac{148150308874659658602710810997156281\ln(7)}{156867260462194671747072000} + \frac{92593017885836069697804708140551751\ln(11)}{165164371759368604798156800}$\\
&$\left.
 + \frac{54895885054227544998269423808649861\ln(13)}{38394084728509185392640000}\right) $\\
&$ +e_t^{28} \left(\frac{4156697661042543089926761131639111\ln(2)}{364802613914495877120000} - \frac{1044633022572944559115574427134187\ln(3)}{1974267456973378158592000} - \frac{1465780771316334046423435211181640625\ln(5)}{593490642521997853241376768}\right.$\\
&$ + \frac{35248751190563249427275944149780968833\ln(7)}{8157097544034122930847744000} - \frac{5514342655611213554403634603472438219\ln(11)}{7707670682103868223913984000}$\\
&$\left. - \frac{323797401611765747818551209929538026763\ln(13)}{82777646674665803706531840000}\right) +O(e_t^{30})$  \\
\end{tabular}
\end{ruledtabular}
\end{table*}

\end{document}